\documentclass[Journal,InsideFigs,SingleSpace, NoLists,SectionNumbers,NoLineNumbers]{ascelike-new}
\pdfoutput=1
\usepackage{amsmath}
\usepackage{float}
\usepackage{multirow}
\usepackage{epsfig}
\usepackage{longtable}
\usepackage{tabu}
\usepackage{graphicx}
\usepackage{subfigure}
\usepackage{amsfonts}
\usepackage{amssymb}
\usepackage{amsbsy}
\usepackage{times}
\usepackage[utf8]{inputenc}
\usepackage{nomencl}
\makenomenclature
\usepackage{mathtools}
\usepackage{overpic}
\usepackage{color,soul}
\usepackage[dvipsnames]{xcolor}
\usepackage{changepage}
\DeclarePairedDelimiter\abs{\lvert}{\rvert}
\makeatletter
\let\oldabs\abs
\def\abs{\@ifstar{\oldabs}{\oldabs*}}
\newcommand{\pd}[2]{\frac{\partial #1}{\partial #2}}
\newcommand{\td}[2]{\frac{d #1}{d #2}}

\newcommand{\ands}{ \ \ \ \mbox{ and } \ \ \ }

\NameTag{Zolfaghari, \today}
\usepackage[toc,page]{appendix}
\begin{document}
		\title{Blade-shaped (PKN) Hydraulic Fracture Driven By A Turbulent Fluid In An Impermeable Rock}	
		
		\author[1]{Navid Zolfaghari}
		\author[2]{Colin R. Meyer}
		\author[1,3]{Andrew P. Bunger}
		
		\affil[1]{Department of Civil and Environment Engineering,
			Swanson School of Engineering, University of Pittsburgh, 
			Pittsburgh, PA  15260, USA.}
		\affil[2]{John A. Paulson School of Engineering and Applied Science,
			Harvard University, 
			Cambridge, MA 02138, USA.}
		\affil[3]{Department of Chemical and Petroleum Engineering,
			Swanson School of Engineering, University of Pittsburgh, 
			Pittsburgh, PA  15260, USA. E-mail: bunger@pitt.edu.}
		\maketitle
		
		\begin{abstract}
			High flow rate, water-driven hydraulic fractures are more common now than ever in the oil and gas industry. Although the fractures are small, the high injection rate and low viscosity of the water, lead to high Reynolds numbers and potentially turbulence in the fracture. Here we present a semi-analytical solution for a blade-shaped (PKN) geometry hydraulic fracture driven by a turbulent fluid in the limit of zero fluid leak-off to the formation. 
			We model the turbulence in the PKN fracture using the Gaukler-Manning-Strickler parametrization, which relates the the flow rate of the water to the pressure gradient along the fracture. The key parameter in this relation is the Darcy-Weisbach friction factor for the roughness of the crack wall. Coupling this turbulence parametrization with conservation of mass allows us to write a nonlinear pde for the crack width as a function of space and time. By way of a similarity ansatz, we obtain a semi-analytical solution using an orthogonal polynomial series. Embedding the asymptotic behavior near the fracture tip into the polynomial series, we find very rapid convergence: a suitably accurate solution is obtained with two terms of the series. This closed-form solution facilitates clear comparisons between the results and parameters for laminar and turbulent hydraulic fractures. In particular, it resolves one of the well known problems whereby calibration of models to data has difficulty simultaneously matching the hydraulic fracture length and wellbore pressure.
		\end{abstract}
	   \section{Introduction}
		Hydraulic fracturing is a method of stimulating relatively impermeable subsurface reservoir rocks to extract oil and gas. In the past two decades, there has been a transition from using high viscosity gels to the use of water in hydraulic fracturing \cite{king2010}. Associated with this change is a 2 to 3 orders of magnitude increase in the characteristic Reynolds number $Re^{*}$, which we define as
			\begin{equation}
			Re^*=\frac{\rho q_{in}}{H\mu},\label{ReynoldsNumber}
			\end{equation}
			where $\rho$ is the fluid density, $q_{in}$ is the volumetric injection rate, $H$ is the hydraulic fracture height $H$, and $\mu$ is the fluid viscosity \cite{tsai2010model}. Table \ref{Re1} shows that a shift from typical gel-based fluids to water leads to an increase from $Re^*\approx$ $0.01$ $-10$ to $Re^*\approx$ 10$^{2}-10^4$, respectively. While a local Reynolds number will vary along the fracture and decrease rapidly near the tip, the Reynolds number defined in Eq. \eqref{ReynoldsNumber} is a constant set by external parameters. In this paper, we describe parameter regimes where $Re$ is large enough for turbulence to exist throughout the hydraulic fracture save the very tip.
			
			\begin{table}
				\centering
				\begin{tabular}{lccccc}
					\hline
					Fluid & Density $\rho$ (kg m$^{-3}$) & Viscosity $\mu$ (Pa$\cdot$s)&  Reynolds Number $Re^*$ \\
					Water & 1000 & 10$^{-3}$ & 10$^{2}$-10$^{4}$\\
					Gel & 1200 & 0.5-1 &  0.01-10\\
					\hline
				\end{tabular}
				\caption{Typical Reynolds numbers for water and gel working fluids with flow rate $q_{in}=0.05-0.2$ m$^{3}$ s$^{-1}$ and fracture height $H = 50-200$ m. }
				\label{Re1}
			\end{table}
			The emerging importance of the turbulent flow regimes will likely increase the number of hydraulic fracture numerical simulations that incorporate turbulence. In order to benchmark these numerical simulations, analytical or semi-analytical solutions are required. One example analytical solution is the large leakoff limit for a PKN hydraulic fracture with rough-walled turbulent flow~\cite{Kano15}. Otherwise, the analytical/semi-analytical solutions, necessary for benchmarking numerical simulations and which do not yet exist for turbulent flow. This is in contrast to a large body of benchmark solutions for the laminar regime (e.g.~\citeNP{geertsma1969rapid,Nordgren72,savitski2002}).
			
			\par
\begin{figure}[H]
				\centering
				\includegraphics[height=70mm,width=70mm]{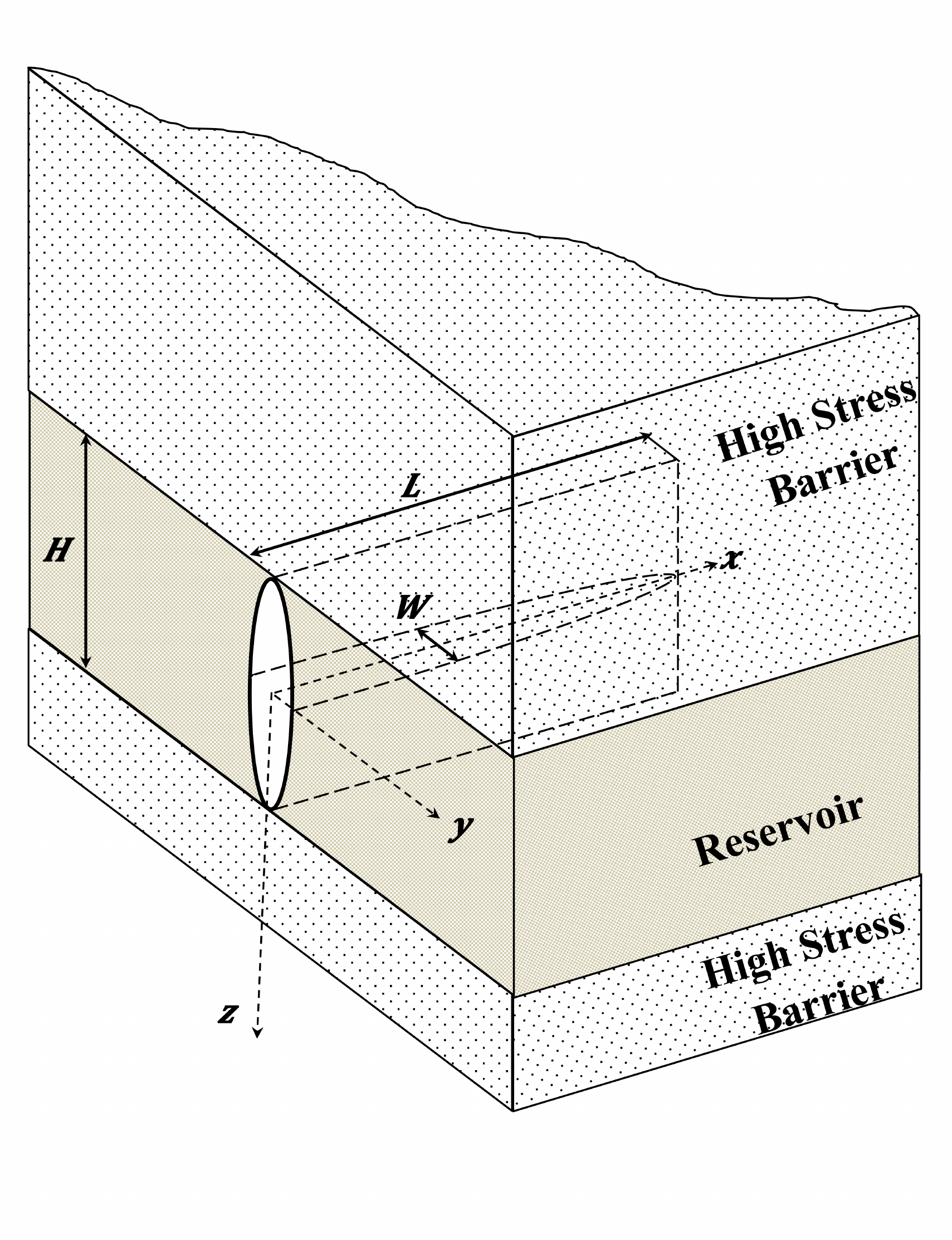}
				\caption{PKN fracture geometry}
				\label{Figure 1}
			\end{figure}

			The tractability of the problem for analytical and semi-analytical solutions requires simple geometries such as plane strain (\citeNP{geertsma1969rapid}), radial (e.g. \citeNP{savitski2002}), and blade-shaped (after \citeNP{perkins61}, and \citeNP{Nordgren72}). While all have usefulness as approximations under particular conditions, the blade-shaped geometry is of practical importance. The minimum stress in most reservoirs is horizontally-directed, leading to vertically-oriented hydraulic fractures. Furthermore, reservoir layers are often bounded by layers that serve to block upward and downward growth of hydraulic fracture, meaning that the horizontal propagation velocity far exceeds the vertical propagation velocity. A large body of field data indicates the resulting blade-like geometry occurs in a wide-range of reservoirs, probably comprising the idealization of the most common fracture geometry (see e.g. \citeNP{de2015s}). The limiting end-member of zero vertical (e.g. height) growth corresponds to the geometry of~\citeNP{perkins61} which was revisited by \citeNP{Nordgren72}, and this so-called ``PKN'' geometry is used in the present study (see Fig.~\ref{Figure 1}). 
			
			The need to consider the turbulent regime for water-driven hydraulic fractures was recognized by~\citeNP{perkins61}. A small number of papers have since considered the turbulent regime of hydraulic fracturing, and some hydraulic fracturing design models (e.g. Meyer 1989) incorporate ability to simulate flow under turbulent conditions. Nilson~(\citeyearNP{Nilson1981gas,Nilson1988gas}) considered plane strain, gas-driven hydraulic fractures under a constant pressure inlet boundary condition. Nilson showed the system evolving among laminar, turbulent, and inviscid regimes and solved the self-similar problems associated with each of these limits of the system. Similarly \citeNP{Emerman1986} examined the problem of a plane strain fluid-driven crack, but instead assuming a constant influx boundary condition. These authors presented an approximate solution, arguing for its practical suitability for modeling magmatic intrusions and natural hydrothermal injections. Turbulent flow is also considered in other geosciences-inspired models. These include the model of drainage of glacial lakes via subglacial fluid-driven cracks developed by \citeNP{tsai2010model}, as well as the model of dyke ascent and propagation developed by \citeNP{Lister1990} and \citeNP{ListerKerr1991}. \citeNP{tsai2010model} used Gaukler-Manning-Strickler (GMS)~\cite{gauckler,Manning1891,strickler23} approximation in order to model turbulent flow for glacial and sub-surface HF. Also \citeNP{tsai2012model} used GMS approach to model near surface hydraulic fracture, motivated by the phenomenon of rapid subglacial drainage. Expanding to account for time-dependent deformation of ice, \citeNP{tsai2015model} used creep flow with GMS to model the rapid glacial lake drainage. These contributions provide a useful background for the fluid flow model, but the boundary conditions and elasticity formulation are specific to their problem and not applicable to industrial HFs. 
			
			More recently \citeNP{Anthonyrajah2013} considered turbulent flow for hydraulic fractures  with blade-shaped geometry with a generalized inlet condition. These authors present analytical solutions for the particular (arguably non-physically motivated) cases of constant fracture speed and volume, and demonstrate a numerical solution method for general injection boundary conditions. The specific case of constant injection rate for a blade-shaped hydraulic fracture was subsequently considered by \citeNP{zia2016laminar}, who point out that many practical cases will consist of flow in the transition between laminar and turbulent flow. For rough-walled fractures, this work numerically demonstrates departure from the laminar solution of about 10-20\% for $Re=2500$ and 30-50\% for $Re=10^5$, as well as complete convergence to a fully turbulent asymptotic solution for $Re=10000$ (though the details of the asymptotic solution are not presented). \citeNP{Dontsov2016}, then, points out the necessity in many practical cases to consider that flow can be in the laminar regime near the tip of the hydraulic fracture and turbulent regime away from the tip. This work is mainly motivated by development of numerical simulations for generalized geometries wherein the behavior near the fracture tip must be properly treated. The solution presented is therefore specified to the moving tip region, allowing the length of the laminar region to be determined and an appropriate tip conditions to be imposed in simulations. 
			
			\par
			In this paper, we present a solution for a PKN-geometry hydraulic fracture driven by a turbulent fluid through an impermeable rock. Although turbulent flows in general remain difficult to describe mathematically, many parameterizations have been developed to describe turbulence through channels and narrow slits. Here we use the solution begins with a generalized expression of the Darcy-Weisbach friction factor, after Gaukler-Manning-Strickler (GMS)~\cite{gauckler,Manning1891,strickler23} approximation for rough-walled turbulence in channel flow. While our solution remains general enough to capture future advances in modeling turbulent flow within hydraulic fractures, provided these can be captured by a power-law relationship between the friction factor and the scale of the fracture roughness. 
			
			The solution presented here is semi-analytical, derived using a Jacobi polynomial series. It follows in the spirit of previous semi-analytical solutions that obtain very rapid series convergence by constructing the family of polynomials so as to embed the appropriate asymptotic behavior near the leading edge~\cite{adachidetornay2002,savitski2002,bungerdetornay2007}. For this reason, our solution method begins with derivation of the near-tip behavior, after which the form of the Jacobi polynomial series is specified. Coefficients of the series are then selected to minimize an objective function that embodies the error of the solution. A convergence study shows that a practically-useful solution is by first two terms of the series. This rapid convergence of the solution justifies the approach, allowing the solution to be written down once for all cases rather than requiring computation for each individual combination of parameters, as is the case for numerical simulations. Finally, the paper concludes with an exploration of the sensitivity to the particulars of the expression for the Darcy-Weisbach friction factor~\cite{weisbach1855,Darcy1857} and with a comparison between solutions resulting from models that impose laminar versus turbulent flow.
			
			\section{Model}\label{sec:types_paper}
			Here we consider a reservoir layer of uniform thickness $H$ contained at the top and bottom by two, higher stress layers assumed to be effective barriers to upward and/or downward hydraulic fracture (HF) growth \cite{perkins61}. A sketch of this geometry is given by Fig.~\ref{Figure 1}. Provided the HF length is several times greater than the thickness $H$, we assume a uniformly-pressurized HF cross section and slowly-varying HF width (opening) with respect to coordinate $x$~\cite{perkins61,adachi2008asymptotic}. These assumptions allow us to derive an expression for the opening $W$ of an elliptical crack in an elastic rock~\cite{Nordgren72}, which is given by
			\begin{equation}\label{elasticity_eq_in}
			W(x,z,t)=\frac{1-\nu}{G}(H^2-4z^2)^{1/2}(p(x,t)-\sigma), 
			\end{equation}
			where $G$ is the shear modulus of elasticity, $\nu$ is the Poisson's ratio, $H$ is the height of the HF. The fluid pressure $p(x,t)$ is taken as uniform in each vertical cross section, and $\sigma$ is the uniform in-situ stress opposing the HF opening.
			\par
			Mass conservation for the incompressible fluid flow in the crack is
			\begin{equation}\label{continuity_eq}
			\frac{\partial A}{\partial t}+\frac{\partial q}{\partial x}=0,
			\end{equation}
			where $A = \pi \omega(x,t)H/4$ is the area of the elliptical crack and $\omega(x,t)$ is the maximum opening in the cross section $\omega(x,t)=W(x,0,t)$. Instead of using the Poiseuille equation for laminar flow~\cite{Nordgren72}, we model the turbulent flow in the crack using the Gaukler-Manning-Strickler \cite{gauckler,Manning1891,strickler23} parametrization
			\begin{equation}
			q_{2D}=\left(-\frac{4W^3}{\rho f_p}\frac{\partial p}{\partial x}\right)^{1/2},
			\label{General_Manning_strickler} 
			\end{equation}
			where $f_p$ is the Darcy-Weisbach friction factor which can be expressed as 
			\begin{equation}
			f_p=m\left(\frac{k}{W}\right)^{\alpha},
			\label{Weisbach}
			\end{equation}
			where $k$ is surface roughness, and $\alpha$ and $m$ are constants with typical values $\alpha=1/3$ and $m=0.143$ \cite{gioia2006,tsai2010model} and we explore the effect of varying these parameters later in the paper. The subscript ``2D" in Eq. \ref{General_Manning_strickler} indicates that the flux is through two-dimensional, horizontal slices at every height $z$. The total flux $q$ is given by the integral over the height of the crack as
			\begin{equation}\label{integralflow}
			q = \int_{-H/2}^{H/2}{q_{2D}dz,}
			\end{equation}
			where the details of this integration are given Appendix~\ref{q_simplified_app}. The result for the total flux is
			\begin{equation}\label{q_simplified}
			\begin{split}
			q &=\Lambda\Upsilon\omega^{2\varphi}\left(-\frac{\partial \omega}{\partial x}\right)^{1/2}, \\
			\Lambda &=\left(\frac{4}{m}\right)^{1/2}B\left(\frac{1}{2},\varphi+1\right),\\
			\Upsilon &=\sqrt{\frac{GH}{4\rho k^{\alpha}(1-\nu)}}, \\
			\varphi & = \frac{3+\alpha}{4},
			\end{split} 
			\end{equation}
			where $B$ is the Beta function (see Eq.\ref{beta_func_orig}, or~\citeNP{Abbramovitz64}), and $\Lambda$ and $\Upsilon$ are parameters of geometry and material properties, respectively. We note that $\Upsilon$ depends on rock properties and reservoir geometry while $\Lambda$ depends only on the parameters of the friction factor $\alpha$ and $m$, which gives a typical value of $\Lambda=7.406$.

			We now substitute the total flux, Eq.~\ref{q_simplified} into the continuity Eq.~\ref{continuity_eq} and define 
			\begin{equation}\label{variablexi}
			\Xi = \frac{4\Lambda \Upsilon}{\pi H}.
			\end{equation}
			Thus, we find the non-linear partial differential equation governing the maximum opening $\omega(x,t)$, which is given by
			\begin{equation}\label{initial_before_scaling_continuity}
			\frac{\partial\omega}{\partial t}=-\Xi\frac{\partial}{\partial x}\left[\omega^{2\varphi}\left(-\frac{\partial\omega}{\partial x}\right)^{\frac{1}{2}}\right].
			\end{equation}
			We then specify three boundary conditions and an initial condition. The third boundary condition is necessary as the total length of the crack $\ell(t)$ is unknown \emph{a priori} and must be determined as part of the analysis. We apply a zero opening initial condition at $t=0$ and by the following boundary conditions:
			\begin{enumerate}
				\item No opening at the crack tip:
				\begin{equation*}
				x=\ell~~\Rightarrow~~\omega(\ell,t)=0.
				\end{equation*}
				\item No fluid loss through the crack tip:
				\begin{equation*}
				x=\ell~~\Rightarrow~~q(\ell,t)=0.
				\end{equation*}
				\item Constant volume rate of flow at the inlet:
				\begin{equation*}
				x=0~~\Rightarrow~~q(0,t)=q_{in}.
				\end{equation*}
			\end{enumerate} 
			\begin{adjustwidth}{0.08\textwidth}{}
				where $q_{in}$ is half of the total fluid injection in the case of symmetric (bi-wing) growth. Alternatively, $q_{in}$ can be the entire injection rate if HF propagation is biased in one direction so as to form a single-wing geometry. As observed in analysis of some field data like Cotton Valley~\cite{rutledge2003}, west Texas~\cite{fischer2008}, east Texas~\cite{mayerhofer2000}, Mound site in Oklahoma~\cite{warpinski1999}, the Lost hill field~\cite{emanuele1998}, and in Barnett shale~\cite{maxwell2002} and also discussed in \cite{wright1999,murdoch2002}.
			\end{adjustwidth}

			\section{Scaling}
			We now look for a similarity solution to Eq. \ref{initial_before_scaling_continuity}. An alternative method of scaling, in the spirit of \citeNP{savitski2002}, is detailed in the Supplementary Materials. This nonlinear pde resembles the equations for viscous gravity currents \cite{Huppert1982} and bouyant hydraulic fractures \cite{Lister1990} and, therefore, we look for a similarity solution of the first kind \cite{Barenblatt1996}. We start by writing the inlet flux as
			\begin{equation}\label{qinkoli}
			q_{in} = \Lambda \Upsilon \left[ w^{2\varphi} \left(-\pd{w}{x}\right)^{1/2}\right]_{x=0}.
			\end{equation}
			Thus, we can scale the pde, Eq.~\eqref{initial_before_scaling_continuity}, and inlet flux conditions as
			\begin{equation}\label{scalingkoli}
			\begin{split}
			\frac{w}{t} &\sim \Xi \frac{w^{2\varphi+1/2}}{x^{3/2}},\\
			\frac{q_{in}}{\Lambda \Upsilon} &\equiv Q \sim \frac{w^{2\varphi+1/2}}{x^{1/2}},
			\end{split}
			\end{equation}
			where we define the scaled flux $Q$ as the ratio of the flux in  $q_{in}$ divided by the intrinsic flux scale $\Lambda \Upsilon$. Combining these two scalings allows us to define the similarity variables
			\begin{equation}\label{scalespatial}
			\xi = \frac{x}{Q^{-2} \left(Q^3 \Xi t\right)^{(4\varphi+1)/(4\varphi+2)}},
			\end{equation}
			and
			\begin{equation}
			w = \left(Q^3 \Xi t\right)^{1/(4\varphi+2)} \Omega(\xi),
			\label{crack_opening}
			\end{equation} 
			We define the length of the crack $\ell(t)$ such that $\xi=1$ coincides with the fracture tip, therefore, we have that
			\begin{equation}
			\ell(t)=Q^{-2} \left(Q^3 \Xi t\right)^{(4\varphi+1)/(4\varphi+2)},
			\label{crack_length}
			\end{equation}
			We can insert the similarity formulation into the governing pde and find
			\begin{equation}
			\Omega-(4\varphi+1)\xi \Omega' = - (4\varphi+2)\left[ \Omega^{2\varphi}\left(- \Omega' \right)^{1/2}\right]',
			\label{ode_omega}
			\end{equation}
			The boundary conditions to Eq.~\eqref{ode_omega} are
			\begin{equation}\label{scaledBCi}
			\begin{split}
			\Omega(\xi=1) =0,\\
			\left[ \Omega^{2\varphi}\left(- \Omega' \right)^{1/2}\right]_{\xi=1}=0,\\
			\left[ \Omega^{2\varphi}\left(- \Omega' \right)^{1/2}\right]_{\xi=0}=1.
			\end{split}
			\end{equation}
			
			\noindent This ode can be integrated once by replacing $\xi \Omega'$ by $(\xi \Omega)'-\Omega$, which gives
			\begin{equation}
			\int_{\xi}^{1}{\Omega d\zeta}-\frac{4\varphi+1}{4\varphi+2}\left[\zeta \Omega\right]_{\xi}^{1} = -\left[ \Omega^{2\varphi}\left(- \Omega' \right)^{1/2}\right]_{\xi}^{1}.
			\end{equation}
			Using the tip boundary conditions, we find
			\begin{equation}\label{int_omega}
			\int_{\xi}^{1}{\Omega~d\zeta}+\frac{4\varphi+1}{4\varphi+2} \xi \Omega = \Omega^{2\varphi}\left(-\Omega' \right)^{1/2}.
			\end{equation}
			To solve this equation, we will use an orthogonal polynomial series method to obtain a semi-analytical solution to Eq. \ref{int_omega}. The opening width changes most rapidly near the tip and, therefore, by embedding the asymptotic solution near the tip, we can derive a rapidly-converging series \cite{savitski2002}. Near the fracture tip, we expect the fracture width $\Omega$ to be small but changing rapidly. Thus, we expect a dominant balance between the second two terms in Eq.~\eqref{ode_omega}, which is equivalent to saying that the integral over the fracture width in the integrated ode, Eq.~\ref{int_omega}, is very small. Thus, to leading order, the near tip behavior is characterized by the ode
			\begin{equation}
			\frac{4\varphi+1}{4\varphi+2} \Omega = \Omega^{2\varphi}\left(-\Omega' \right)^{1/2}.
			\end{equation}
			Simplifying and separating, we find
			\begin{equation}
			-\int_{\xi}^{1}{\Omega^{4\varphi-2}~d\Omega}=\left(\frac{4\varphi+1}{4\varphi+2}\right)^{2}\left( 1-\xi\right).
			\end{equation}
			Thus, the solution for the width near the tip is
			\begin{equation}
			\Omega=\left[\sqrt{4\varphi-1}\left(\frac{4\varphi+1}{4\varphi+2}\right)\right]^{2/(4\varphi-1)}\left(1-\xi\right)^{1/(4\varphi-1)}.
			\label{TipAsymptotics}
			\end{equation}
			We analyze the tip region later in the paper and show that, although, there is relaminarization in a small boundary layer near the tip it is a sufficiently small region that the turbulent expression derived in Eq.~\ref{TipAsymptotics} still holds. 
			
			Embedding this asymptotic solution into the polynomial series allows us to approximate the series using only a few terms which clearly shows the dependence upon the parameters and can be readily adopted for benchmarking purposes. Numerical solutions, although certainly feasible with existing methods, would not provide the insights or usability of a semi-analytical solution.
			
			\section{Solution}
			\subsection{Overview of the Method}
			To solve Eq. \ref{int_omega}, we construct an orthogonal polynomial series~\cite{savitski2002}. Orthogonal polynomials are sets of functions that follow
			\begin{equation}\label{R(c)}
			\int_a^b R(x) B_m(x) B_n(x)dx=0,
			\end{equation}
			for all $m\neq n$ (where $R(x)$ is the weight function) and 
			\begin{equation}
			\int_a^bR(x) B_n(x)^2dx=h_n,
			\end{equation} 
			if $m=n$~\cite{Abbramovitz64}. The proposed solution is thus in the form of infinite series using basis functions $\hat{\Omega}_{k}$
			\begin{equation}\label{series_exp_infinit_global}
			\Omega=\sum_{i=0}^\infty {\mathcal{A}_i \hat{\Omega}_i}, 
			\end{equation}
			where $\mathcal{A}_i$ are coefficients selected so that the solution satisfies the governing equations. The base functions $\hat{\Omega}_i$ must be orthogonal, therefore
			\begin{equation}\label{Orthogonality}
			\int_0^1\hat{\Omega}_i\hat{\Omega}_jdx=\delta_{ij},
			\end{equation}
			where $\delta_{ij}$ is the Kronecker delta function.
			\par
			Rapid convergence of the series is promoted by selecting the base functions so as to embed the near-tip behavior \cite{savitski2002}, which we found to be of the form $\Omega\sim\mathcal{X}(1-\xi)^{\mathcal{B}}$ where $\mathcal{X}$ and $\mathcal{B}$ are
			\begin{equation}
			\mathcal{X}=\left[\sqrt{4\varphi-1}\left(\frac{4\varphi+1}{4\varphi+2}\right)\right]^{2/(4\varphi-1)} \ands \mathcal{B}=\frac{1}{4\varphi-1}.
			\label{int_omega_solved}
			\end{equation}
			The base functions will then be constructed so that 
			\begin{equation}
			\hat{\Omega}_i=\mathfrak{D}_i\mathfrak{f}_i(\xi)\mathcal{X}(1-\xi)^{\mathcal{B}},
			\end{equation}
			where $\mathfrak{D}_i$ are constants chosen so as to satisfy the orthogonality relationship, Eq.~\ref{Orthogonality}. Upon substitution
			\begin{equation}\label{Orthogonalitytwo}
			\int_0^1(\mathfrak{D}_i\mathfrak{D}_j)\mathcal{X}^2(1-\xi)^{2\mathcal{B}} \mathfrak{f}_i(\xi)\mathfrak{f}_j(\xi)d\xi=\delta_{ij}.
			\end{equation}
			A convenient choice for the functions $\mathfrak{f}_i$ are the Jacobi polynomials, which have the following orthogonality relationship \cite{Abbramovitz64}
			\begin{equation}\label{Integral_h_function}
			\int_{0}^{1}(1-\xi)^{c-e}\xi^{e-1}G_i(c,e,\xi)G_j(c,e,\xi)d\xi=h_i(c,e)\delta_{ij},
			\end{equation}
			where $G_i(c,e,\xi)$ is the $i^{th}$ order Jacobi polynomial, expressible as   
			\begin{equation}\label{G_function}
			G_i(c,e,\xi)=\frac{\Gamma(e+i)}{\Gamma(c+2i)}\sum_{j=0}^{i}(-1)^j\binom{i}{j}\frac{\Gamma(c+2i-j)}{\Gamma(e+i-j)}\xi^{i-j},
			\end{equation}
			where $\Gamma(i)$ is the Gamma function~\cite{Abbramovitz64} and $h_i(c,e)$ is the norm of $G_i(c,e,\xi)$ and given by
			\begin{equation}\label{h_function}
			h_i(c,e)=\frac{i!\Gamma(i+e)\Gamma(i+c)\Gamma(i+c-e+1)}{(2i+c)\Gamma^2(2i+c)}.
			\end{equation}
			Setting $e=1$, $c=2\mathcal{B}+1$ and rearranging Eq.~\ref{Integral_h_function} we have
			\begin{equation}\label{Integral_h_functiontwo}
			\int_{0}^{1}\left(\frac{1}{h_i\left(2\mathcal{B}+1,1\right)}\right)(1-\xi)^{2\mathcal{B}}G_i\left(2\mathcal{B}+1,1,\xi\right)G_j\left(2\mathcal{B}+1,1,\xi\right)d\xi=\delta_{ij}.
			\end{equation}
			Now comparing Eq.~\ref{Integral_h_functiontwo} with Eq.~\ref{Orthogonalitytwo} leads to the conditions
			\begin{equation}\label{Integral_h_functionanswer}
			\begin{split}
			\mathcal{D}_i &=\frac{1}{\mathcal{X}h_i^{1/2}\left(2\mathcal{B}+1,1\right)},\\
			\mathfrak{f}_i(\xi) &=G_i\left(2\mathcal{B}+1,1,\xi\right).
			\end{split}
			\end{equation}
			As a result, the base functions are given by
			\begin{equation}\label{jacobi_setPrimary}
			\hat{\Omega}_i=\frac{(1-\xi)^\mathcal{B}}{h_i^{1/2}\left(2\mathcal{B}+1,1\right)}G_i\left(2\mathcal{B}+1,1,\xi\right).
			\end{equation}
			Using the asymptotic near-tip solution, the form of the orthogonal base functions is therefore given by
			\begin{equation}\label{jacobi_set}
			\hat{\Omega}_i=\frac{(1-\xi)^{1/(4\varphi-1)}}{\sqrt{h_i\left(\frac{4\varphi+1}{4\varphi-1},1\right)}}G_i\left(\frac{4\varphi+1}{4\varphi-1},1,\xi\right),
			\end{equation}
			where $h_i$ is the norm of the basis.

				\subsection{Calculating Coefficients of the Series}
				Given the basis functions in Eq. \ref{jacobi_set}, we can now calculate the coefficients $\mathcal{A}_i$ of the series in Eq. \ref{series_exp_infinit_global}. The approach is to designate equally spaced control points on $0<\xi<1$ (we typically used 10 control points) and retain the first $n$ terms of the polynomial series. 
				
				Now, it is possible to construct a residual function in terms of $\mathcal{A}_i$ and minimize that function. The chosen residual function embodies the sum of the squares of the mismatch between the left and right hand sides Eq. \ref{int_omega} at each control point~\cite{savitski2002}. Its formal expression is given by
				\begin{equation}\label{residual_function}
				\Delta(\mathcal{A}_1,\cdots,\mathcal{A}_n)=\sum\limits_{i=1}^{\mathcal{Q}}{\left(\frac{\Delta^L(\xi_i,\mathcal{A}_1,\cdots,\mathcal{A}_n)}{\Delta^R(\xi_i,\mathcal{A}_1,\cdots,\mathcal{A}_n)}-1\right)^2},
				\end{equation}
				where $\Delta^L(\xi_i,\mathcal{A}_1,\cdots,\mathcal{A}_n)$ is the left side of Eq.~\ref{int_omega} for specific value of $\xi_i$ and $\Delta^R(\xi_i,\mathcal{A}_1, \cdots, \mathcal{A}_n)$ is similarly the right side of Eq.~\ref{int_omega} with the solution parametric in the flow law parameter $\varphi$. By minimizing $\Delta(\mathcal{A}_1, \cdots, \mathcal{A}_n)$, we can find the value of each unknown variable $\mathcal{A}_i$. 
				\subsection{Length}
				Now we express evolution of the crack length with respect to time. Recalling Eq.~\ref{crack_length}, 
				\[\ell(t)=Q^{-2} \left(Q^3 \Xi t\right)^{(4\varphi+1)/(4\varphi+2)}.\]
				the crack length evolves with time with power of ${(4\varphi+1)/(4\varphi+2)}$. For illustration, if we consider values $\alpha=1/3$ and $m=0.143$ for the Darcy-Weisbach friction factor \cite{tsai2010model}, from Eq.~\ref{q_simplified} the value of $\varphi$ will be $5/6$. Therefore, the value of the power in Eq.~\ref{crack_length} for turbulent flow is
				\begin{equation}\label{power_turb}
				\frac{4\varphi+1}{4\varphi+2}=\frac{13}{16},
				\end{equation}
				while for laminar flow this value is $4/5$ (see Eq.~\ref{Nordgren_solution}, recall from \citeNP{Nordgren72} solution for no-leakoff case). By comparing the value of the power of the time in length formula for laminar and turbulent flow, we can see that the power in laminar flow is within 1\% of turbulent flow and they are very close to each other.
				
				\subsection{Pressure}
				
				In order to find the distribution of pressure along the HF, we invoke elasticity (Eq.~\ref{elasticity_eq_z_zero}). The net pressure inside the crack is therefore expressed in terms of the maximum opening at each cross section, the height of the crack $H$, and material properties of the rock according to
				\begin{equation}\label{pressure_total_eq}
				p(x,t)-\sigma\equiv p_{net}(x,t)=\frac{G}{1-\nu}\frac{w(x,t)}{H}.
				\end{equation}
				Thus, by replacing the opening from Eq.~\ref{crack_opening} into Eq.~\ref{pressure_total_eq}
				\begin{equation}\label{pressure_beginner_terms}
				p_{net}(x,t) =\frac{G}{1-\nu}\frac{1}{H}\left(Q^3 \Xi t\right)^{1/(4\varphi+2)} \Omega(\xi).
				\end{equation}
				Now by combining Eq.~\ref{pressure_beginner_terms} and Eq.~\ref{series_exp_infinit_global}
				\begin{equation}\label{pressure_implicite_format}
				p_{net}(x,t) =\frac{G}{1-\nu}\frac{1}{H}\left(Q^3 \Xi t\right)^{1/(4\varphi+2)}\sum_{i=0}^\infty {\mathcal{A}_i\hat{\Omega}_i(\xi(x,t))}.
				\end{equation}
				Hence, given the solution for the series coefficients $\mathcal{A}_i$, Eq.~\ref{jacobi_set}, and the equation for length scale parameter $\xi=x/\ell(t)$, the pressure is readily computed.  
				
				\section{Results}
\begin{figure}[H]
					\centering
					\includegraphics[height=50mm]{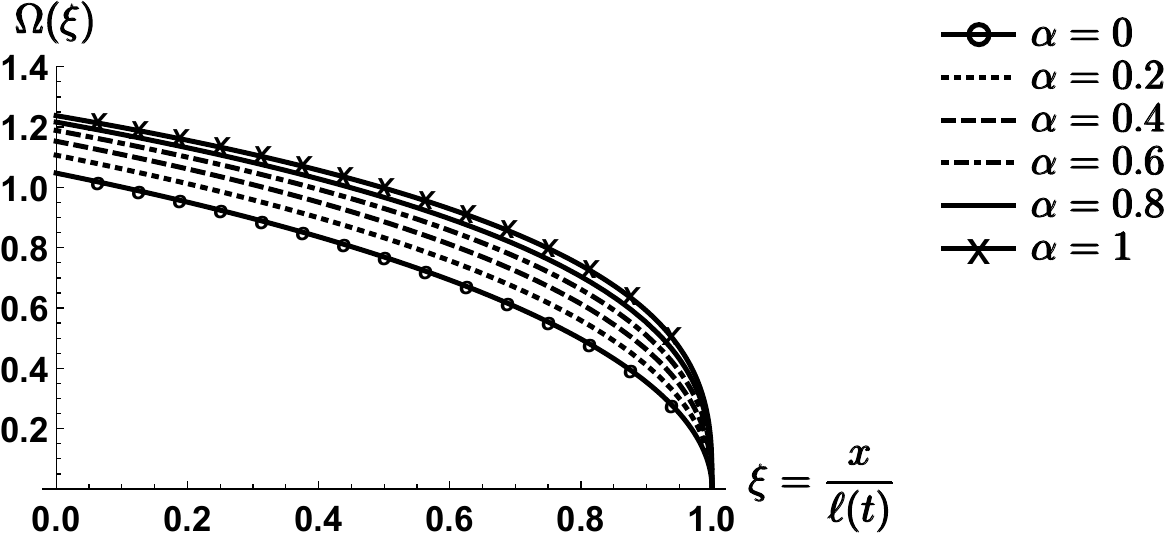}
					\caption{Variation of scaled opening with respect to change of dimensionless length $\xi$ for different value of $\alpha$.}
					\label{Figure 2}
				\end{figure}

\begin{figure}[H]
					\centering
					\includegraphics[height=50mm]{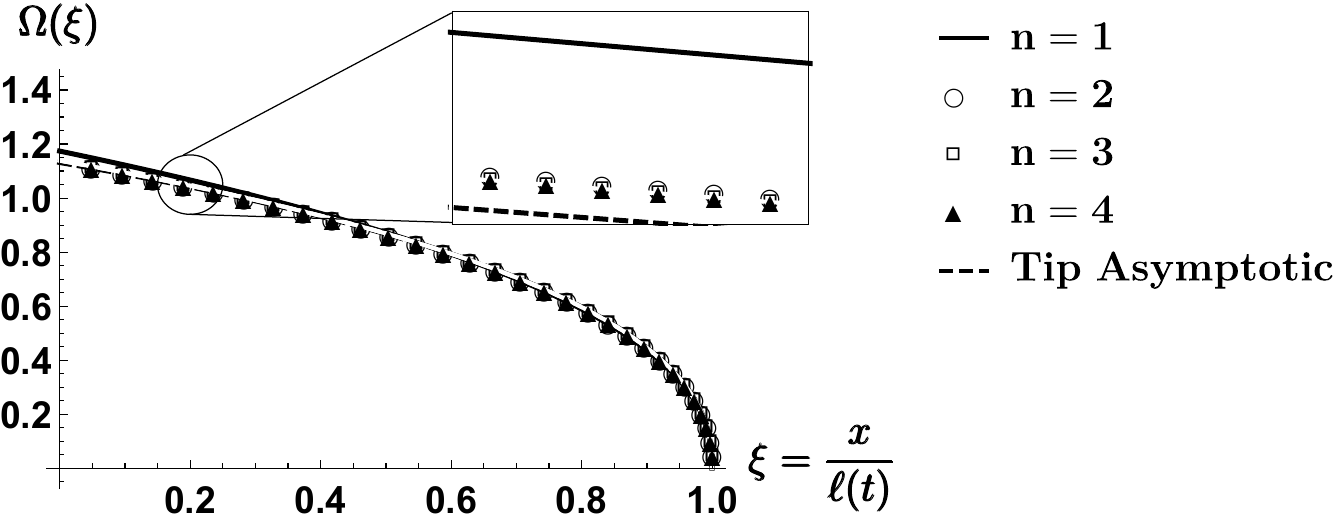}
					\caption{Scaled opening along the hydraulic fracturing. Noting that using just one term is a very good approximation and $n\geq2$ gives solutions that are indistinguishable. Dashed line correspond to tip asymptotic solution (look Supplementary Materials).}
					\label{Figure 4}
				\end{figure}

\begin{figure}[H]
					\centering
					\includegraphics[height=70mm]{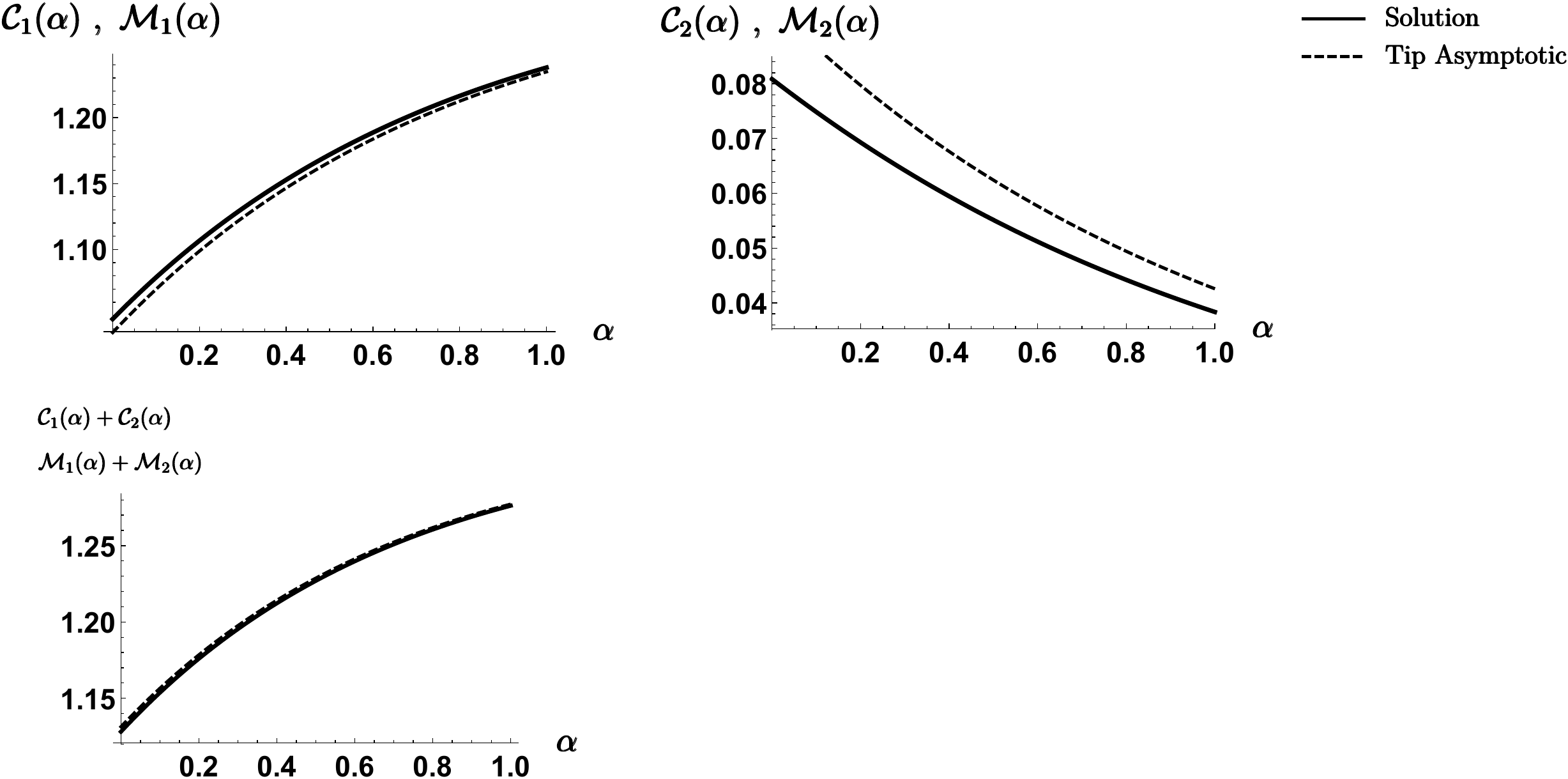}
					\caption{Constants $C_{i}$ and $M_{i}$ in Eqs.~\protect\ref{opening_alpha_different_form} and~\protect\ref{assymtipsol} as functions of $\alpha$.}
					\label{Figure 5}
				\end{figure}

				The dimensionless opening depends only on the exponent $\alpha$ from the Darcy-Weisbach friction factor through $\varphi$ (refer to Eq.~\ref{q_simplified}). Fig.~\ref{Figure 2} shows the sensitivity of the dimensionless opening $\Omega$ with respect to $\alpha$. 
				It is clear from these results the sensitivity of the dimensionless opening, $\Omega$, to $\alpha$ is relatively small. However, the actual opening and length will be strongly affected via the dependence of the scaling quantities on $\alpha$ (see Eq.~\ref{crack_opening} and~\ref{crack_length}).
				Here we consider the particular values $\alpha=1/3$ and $m=0.143$ for the parameters of the Darcy-Weisbach friction factor. We truncate the Jacobi polynomial series at four terms and the coefficients are given in Table~\ref{Table 1}. From Fig.~\ref{Figure 4}, it can be seen that after second term ($n=2$) the solution is indistinguishable with additional terms. This shows the very rapid convergence enabled by embedding the tip asymptotic behavior in the form of the base functions. The truncated solution for $n=2$ terms for opening is given as
				\begin{table}
					\caption{Coefficients for Jacobi Polynomial series for $n=1,2,3,4$ with $\alpha=1/3$.} 
					\centering \vspace{0.25 cm}
					\renewcommand{\arraystretch}{1.5}
					\begin{tabu}{>{\centering}m{1.2cm} >{\centering}m{3cm} >{\centering}m{3cm} >{\centering}m{3cm} >{\centering}m{3cm}}
						\firsthline \hline
						\large\textbf{n} & \large\textbf{1}         & \large\textbf{2}         & \large\textbf{3}         & \large\textbf{4} \\ 
						\hline 
						$\mathcal{A}_1$  & $8.616\times 10^{-1}$    & $8.517\times 10^{-1}$    & $8.515\times 10^{-1}$    & $8.515\times 10^{-1}$  \\ 
						$\mathcal{A}_2$  & -                        & $1.115\times 10^{-2}$    & $1.124\times 10^{-2}$    & $1.124\times 10^{-2}$  \\ 
						$\mathcal{A}_3$  & -                        & -                        & $2.904\times 10^{-4}$    & $2.954\times 10^{-4}$  \\ 
						$\mathcal{A}_4$  & -                        & -                        & -                        & $6.156\times 10^{-6}$  \\     
					\end{tabu} 
					\label{Table 1} 
				\end{table} 
				\begin{equation}\label{explicitFORM}
				\begin{split}
				\Omega &=(1-\xi)^{\frac{3}{7}} (1.1387 + 0.0626\xi),\\
				\lambda &=1.0874.
				\end{split}
				\end{equation}
				The truncated solution in dimensional variables can be obtained from similarity scaling, Eq.~\ref{explicitFORM}, which gives
				\begin{equation}\label{ImplicitFORMall}
				\begin{split}
				w(x,t) &=0.8122~\left(\frac{k^{1/3}q_{in}^3\rho}{H^2}\frac{1-\nu}{G}\right)^{\frac{3}{16}}t^{\frac{3}{16}}\left(1-\frac{x}{\ell(t)}\right)^{\frac{3}{7}}\left(1+0.05497~\frac{x}{\ell(t)}\right),\\
				\ell(t)&=2.1619~\frac{q_{in}}{H}\left(\frac{H^2}{k^{1/3}q_{in}^3\rho}\frac{G}{1-\nu}\right)^{\frac{3}{16}}t^{\frac{13}{16}},\\
				p_{net}(x,t) &=0.8122~\left(\frac{kq_{in}^9\rho^3}{H^{22}}\left(\frac{G}{1-\nu}\right)^{13}\right)^{\frac{1}{16}}t^{\frac{3}{16}}\left(1-~\frac{x}{\ell(t)}\right)^{\frac{3}{7}}\left(1+0.05497~\frac{x}{\ell(t)}\right),\\
				q &=q_{in}\left(1-~\frac{x}{\ell(t)}\right)^{\frac{3}{7}}\left(1+0.05497~\frac{x}{\ell(t)}\right)^{\frac{5}{3}}\left(1+0.21005~\frac{x}{\ell(t)}\right)^{\frac{1}{2}}.
				\end{split}
				\end{equation}
				We can also write this truncated solution for general form of $\Omega$ as  
				\begin{equation}\label{opening_alpha_different_form}
				\Omega =(1-\xi)^{\frac{1}{2+\alpha}} \left(\mathcal{C}_1(\alpha) + \mathcal{C}_2(\alpha)\xi\right),
				\end{equation}
				where $\mathcal{C}_1(\alpha)$ and $\mathcal{C}_2(\alpha)$ are constants that vary with the value of $\alpha$ as shown in Fig.~\ref{Figure 5}.
				
				The rapid convergence of the polynomial series motivates us to derive a second order asymptotic solution near the tip (see Supplementary Materials for derivation of this solution). The asymptotic solution has the same structure as the truncated solution and is given by
				\begin{eqnarray}\label{assymtipsol}
				\begin{aligned}
				\Omega_{tip} &=(1-\xi)^{\frac{1}{2+\alpha}} \left(\mathcal{M}_1(\alpha) + \mathcal{M}_2(\alpha)\xi\right),\\
				\mathcal{M}_1(\alpha) &= (\alpha+2)^{-\frac{\alpha+1}{\alpha+2}} \left(\frac{\alpha+4}{\alpha+5}\right)^{2/(\alpha+2)}  \left(1+\alpha+\frac{(\alpha+2)(\alpha+5)}{(\alpha+3)(\alpha+4)}\right),\\
				\mathcal{M}_2(\alpha) &=(\alpha+2)^{-\frac{\alpha+1}{\alpha+2}} \left(\frac{\alpha+4}{\alpha+5}\right)^{2/(\alpha+2)}  \left(1-\frac{(\alpha+2)(\alpha+5)}{(\alpha+3)(\alpha+4)}\right).
				\end{aligned}
				\end{eqnarray} 
				A comparison between the asymptotic solution and the truncated solution is given in Fig.~\ref{Figure 5}
				
				\subsection{Applications}
				We now present a few examples illustrating the practical relevance of the newly-derived solution. The purpose is:
				\begin{enumerate}
					\item  To provide a comparison between laminar and turbulent flow solutions for an example high-rate water-driven HF to show that the difference is significant and therefore use of the laminar model in instances where Reynolds number is large can lead to substantial errors.
					\item To compare the expected Reynolds numbers for different fluid families (refer to the Supplementary Materials) in order to clarify conditions under which the turbulent and laminar models are expected to be relevant. 
					\item To show the size of the near-tip laminar zone relative to the fracture as a whole, thereby clarifying conditions under which the majority of the HF is in turbulent regime.
				\end{enumerate}
				Here we compare the turbulent solution with the laminar solution of \citeNP{Nordgren72}, where the opening, net pressure, and length are given by
				\begin{equation}\label{Nordgren_solution}
				\begin{split}
				w_N(0,t) &=2.5\left[\frac{\mu q_{in}^2}{H}\frac{(1-\nu)}{G}\right]^{\frac{1}{5}}t^{\frac{1}{5}},\\
				\ell_N(t) &=0.68\left[\frac{q_{in}^3}{\mu H^4}\frac{G}{(1-\nu)}\right]^{\frac{1}{5}}t^{\frac{4}{5}},\\
				p_{net_N}(0,t) &=2.5\left[\frac{\mu q_{in}^2}{H^6}\frac{G^4}{(1-\nu)^4}\right]^{\frac{1}{5}}t^{\frac{1}{5}}.
				\end{split}
				\end{equation}
				
				The parameter values used in the example case are given in Table~\ref{Table 2} and the characteristic Reynolds number is $Re^*=10^4$. At this Reynolds number, the flow is turbulent for all values of roughness and, therefore, the \citeNP{Nordgren72} solution does not apply. Hence, this comparison illustrates the magnitude of the error associated with inappropriately choosing the laminar model instead.
				
\begin{figure}[H]
					\centering
					\includegraphics[height=50mm]{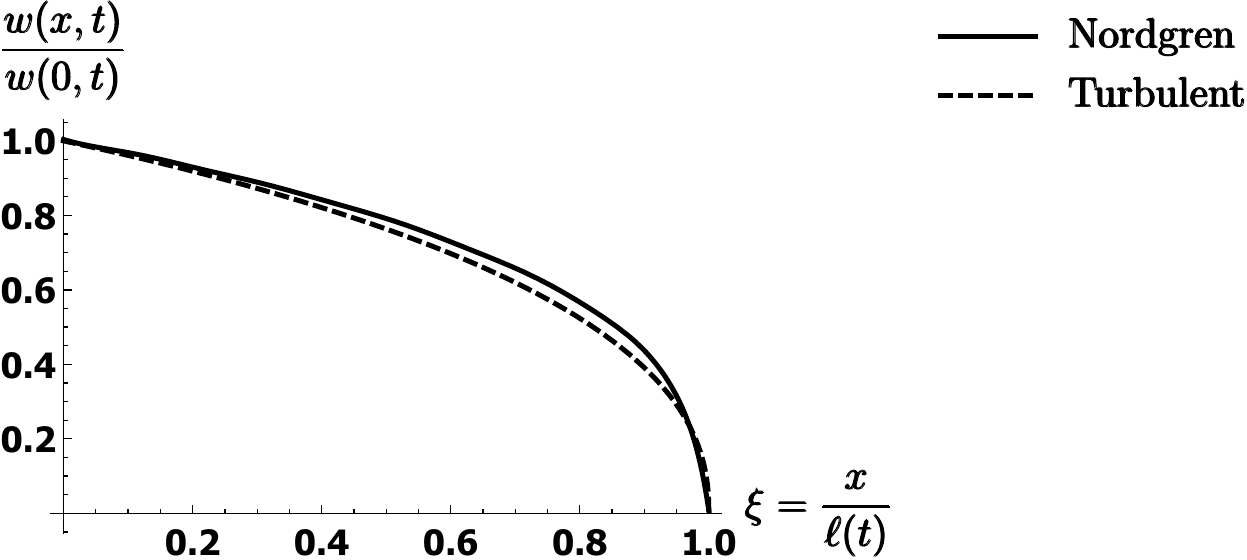}
					\caption{Normalized variation of maximum fracture width in different cross sections for laminar and turbulent flow.}
					\label{Figure 7}
				\end{figure}

\begin{figure}[H]
					\centering
					\begin{overpic}[height=80mm]{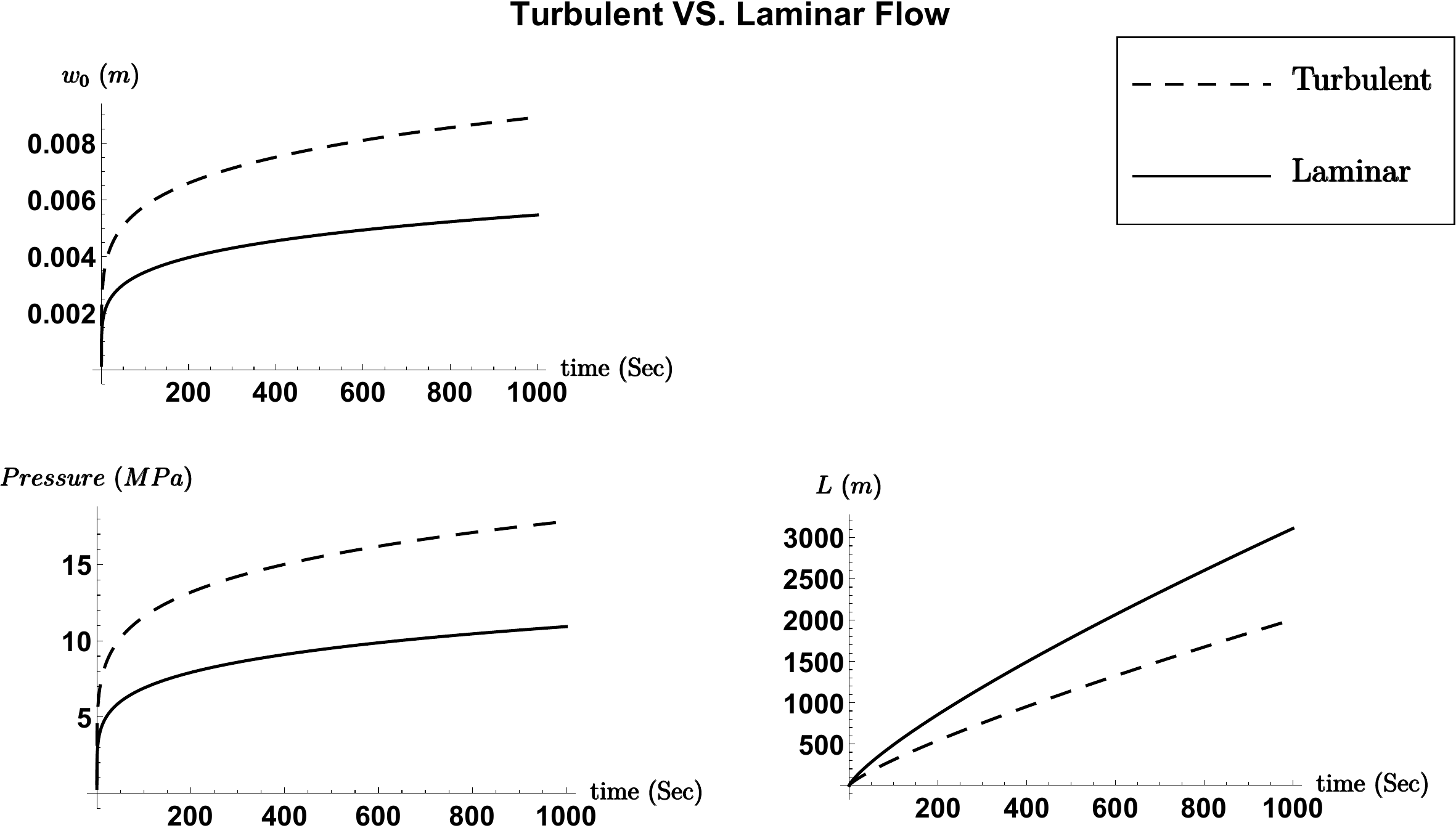}\put(1.5,0){(b)}\put(51,0){(c)}
						\put(1.5,27){(a)}\end{overpic}
					\caption{Comparison between laminar and turbulent solutions for the parameters given in Table~\protect\ref{Table 2}. (a) Maximum fracture width at $x=0$, (b) Predicted fluid net pressure at the wellbore, (c) Fracture length.} 
					\label{Figure 8}
				\end{figure}

				Fig.~\ref{Figure 7} shows that the crack opening profile for the turbulent solution is similar to the \citeNP{Nordgren72} solution for laminar flow when the opening is normalized by the opening at the wellbore. However, the magnitude of the opening, shown in Fig.~\ref{Figure 8}a, is over 50\% greater for the turbulent model. This greater opening is caused by a larger fluid net pressure in the turbulent case (Fig.~\ref{Figure 8}b). Finally, because the total volume is the same in both cases, the laminar model overpredicts the length by over 40\% (Fig.~\ref{Figure 8}c). Hence, the turbulent model shows that high Reynolds number treatments will result in higher pressure, greater widths, and shorter lengths than predicted by incorrectly-applied laminar models. This result is also consistent with comparisons for the large leak-off PKN-type solution presented by \citeNP{Kano15} and discussed based on scaling arguments by \citeNP{AmBu15}.
				
				\begin{table}
					\caption{Material Properties and physical constants for illustration. }
					\centering \vspace{0.25 cm}
					\renewcommand{\arraystretch}{1.5}
					\begin{tabu}{>{\centering}m{2cm} >{\centering}m{2.5cm}}
						\hline\hline
						\textbf{Parameter} & \textbf{Value}                           \\ 
						\hline 
						$q_{in}$   & 0.2~m$^{3}$ s$^{-1}$             \\ 
						$\nu$      & $0.25$                          \\ 
						$\mu$      & $0.001$ Pa$\cdot$s                    \\ 
						$\rho$     & 1000~kg m$^{-3}$     \\ 
						$k$        & 0.3~mm                      \\
						$m$        & $0.143$                         \\
						$H$        & $20$~m                          \\
						$G$        & $30$~GPa                      \\
						$\alpha$   & $1/3$                   \\
					\end{tabu} 
					\label{Table 2} 
				\end{table} 
				
				
				It is therefore shown that substantial errors in predictions can arise due to misuse of the laminar or turbulent models. Instead, the choice should be made based on a calculated value of Reynolds number characterizing the regime for a given case.
				
				To address the second objective of this section, comparing the expected Reynolds numbers for different fluid families, we examine four fluids. To see the effect of changing the fluid on the value of the characteristic Reynolds number, we assume the flux $q_{in}$ and height $H$ are constant and use the values given in Table~\ref{Table 2}. Referring to Table~\ref{Table 3}, the value of $\mu/\rho$ for different fluids is given. The biggest number in the table corresponds to cross-linked gel and is equal to $41.67\times 10^{-5}$ m$^2$ s$^{-1}$. The smallest value is associated with CO$_2$ ($8.34\times 10^{-8}$ m$^2$ s$^{-1}$). Hence, the ratio of characteristic Reynolds number for those two cases is equivalent to the ratio of density over viscosity for those two fluids, namely around 5000. This contrast of Reynolds number can be large enough to change the flow regime from laminar to turbulent.
				
				\begin{table}
					\caption{Different fracturing fluids and their rheology.} 
					\centering \vspace{0.25 cm}
					\renewcommand{\arraystretch}{1.5}
					\begin{tabu}{>{\centering}m{5cm} >{\centering}m{3cm} >{\centering}m{2.5cm} >{\centering}m{4cm}} 
						\hline\hline
						\multirow{2}{*}{Fluid}            & Density    & Viscosity    & Kinematic Viscosity   \\  
						&     $\rho$ (kg m$^{-3}$)        &     $\mu$ (Pa$\cdot$s)    & $\frac{\mu}{\rho}$ (m$^2$ s$^{-1}$)                       \\ 
						\hline 
						Slick-water                       & $1000$               & $0.001$            &  $10^{-6}$           \\ 
						X-linked Gel                      & $1200$               & $0.5$              &   $41.67\times 10^{-5}$       \\ 
						CO$_2$ (Supercritical CO$_2$)    & $600$                & $5\times 10^{-5}$  &  $8.34\times 10^{-8}$       \\ 
						Linear Gel                        & $1200$               & $0.05$             &  $41.67\times 10^{-6}$          \\ 
					\end{tabu} 
					\label{Table 3} 
				\end{table}
				
				We now examine the role of geometry and pumping rate. Typical heights of HFs fall in the range 20 m $<H<200$ m \cite{fisherwarpinski2012}. We will take the range of injection rates from 0.01 m$^3$ s$^{-1}$$<q_{in}<$ 0.2 m$^3$ s$^{-1}$. Hence the ratio is $5\times 10^{-5}$ m$^2$ s$^{-1}$ $<q_{in}/H<$ 0.01 m$^2$ s$^{-1}$. Typical Reynolds numbers for the 4 fluids and their densities and viscosities are listed in Table~\ref{Table 3}. Based on open channel problems~\cite{henderson1966open,munson02}, the corresponding range Reynolds number for the laminar regime $Re<500$, while $Re>12500$ is considered as turbulent regime. The values in between are thus considered to occupy a transition from laminar to turbulent flow. Accordingly, fracturing with CO$_2$ will be mostly turbulent flow. Water in most cases is transition and in the most field relevant cases is closer to, and therefore better approximated by, the turbulent regime. The other two fluids lead to laminar flow (for more details, refer to to the Supplementary Materials).

				The suggested Reynolds numbers are experimental determined and may change based on geometric details for open channel problems. For most of the practical cases, Reynolds number less than 500 is laminar. However, there is no definitive upper limit defining the transition to turbulent flow~\cite{te1959open,munson02,gioia2006}. In open channel problems an upper limit for the transition depends on other parameters like the channel geometry. Therefore, we discuss two alternative methods for estimating an appropriate $Re$ to define transition to turbulent flow in order to select an appropriate fluid flow law. First, we compare the characteristic fluid pressure associated with laminar flow $P_{laminar}$ to the characteristic pressure associated with turbulent flow $P_{turb}$. In this approach we define $P_{laminar}>P_{turb}$ as the laminar regime and $P_{laminar}<P_{turb}$ as the turbulent regime. Although the transition $Re$ depends on the fluid properties (see the Supplementary Materials for details), a typical transition value is around $Re=500$. With this definition and the proposed ranges for different parameters, CO$_2$ is always turbulent and water is turbulent for nearly all relevant cases (see Supplementary Materials for more details).
				
				Returning (briefly) to the definition of GMS, to develop the friction factor, hydraulic radius is used. Hydraulic radius ($R_h$) is a characteristic length that helps to calculate the effect of different cross sections. Originally, this parameter introduced so that pipe flow equations can be expanded to other non-circular conduits. Mathematically, the hydraulic radius is the ratio of cross section of the fluid flow over the wetted perimeter ($R_h=A/\mathfrak{P}$). For elliptical cracks, the area is $A=\pi\omega H/4$, and if the eccentricity of the ellipse defined as $e=\sqrt{1-(\omega/H)^2}\approx 1$ the perimeter of it is defined as
				\begin{equation}
				\mathfrak{P}=\pi H\left[1-\sum_{i=1}^{\infty}\frac{(2i)!^2}{(2^i.i!)^4}\frac{e^{2i}}{2i-1}\right]\approx 2H,
				\end{equation}	
				so the value of hydraulic radius is $\pi \omega/8$. In the PKN model, the cross section is ellipse, so the average value of the opening in one cross section is \cite{zia2016laminar}
				\begin{equation}
				\overline{\omega}=\frac{1}{H}\int_{-H/2}^{H/2}W(x,z,t)dz=\frac{\pi}{4}\omega,
				\end{equation}
				thus, hydraulic radius is half the value of average opening of the crack at specific cross section ($R_h=\overline{\omega}/2$).	
				
				The Moody diagram can be used to determine the transition $Re$ for the purpose of selecting a fluid flow model (see the Supplementary Materials for more details). The premise of this argument is that for most cases in HF the scaled value of the fracture roughness is in order of 0.05 or higher ($k/w>0.05$). According to the Moody diagram (Supplementary Materials), for such a roughness, the fully turbulent regime is occurring at $Re>10^4$ and the transition from laminar starts around $Re>2000$. We further note that the friction factor in this transition regime is for the most part closely enough approximated by the turbulent GMS model that it is a viable selection from a practical perspective.
				
\begin{figure}[H]
					\centering
					\begin{overpic}[height=100mm]{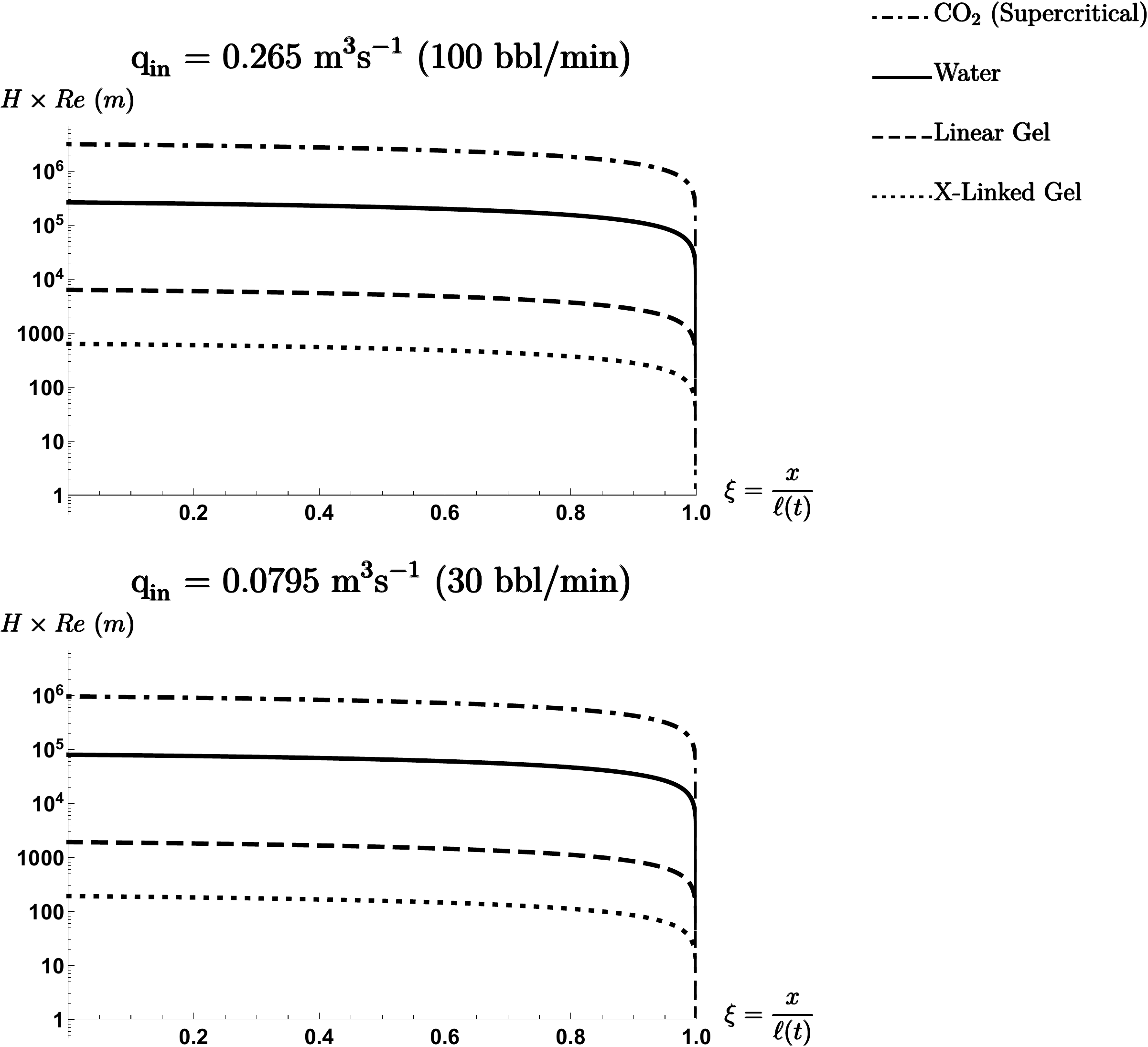}\put(-7,50){(a)}\put(-7,0){(b)}
					\end{overpic}
					\caption{Reynolds number variation along the crack. By knowing the height of the barrier $H$, it is possible to find the Reynolds number at any points inside the crack. The time dependence is embedded in length of the crack which can be seen in $x/\ell(t)$. Note that this graph still depends on pumping rate $q_{in}$.}
					\label{Figure 11}
				\end{figure}

\begin{figure}[H]
					\centering
					\centerline{\includegraphics[height=60mm]{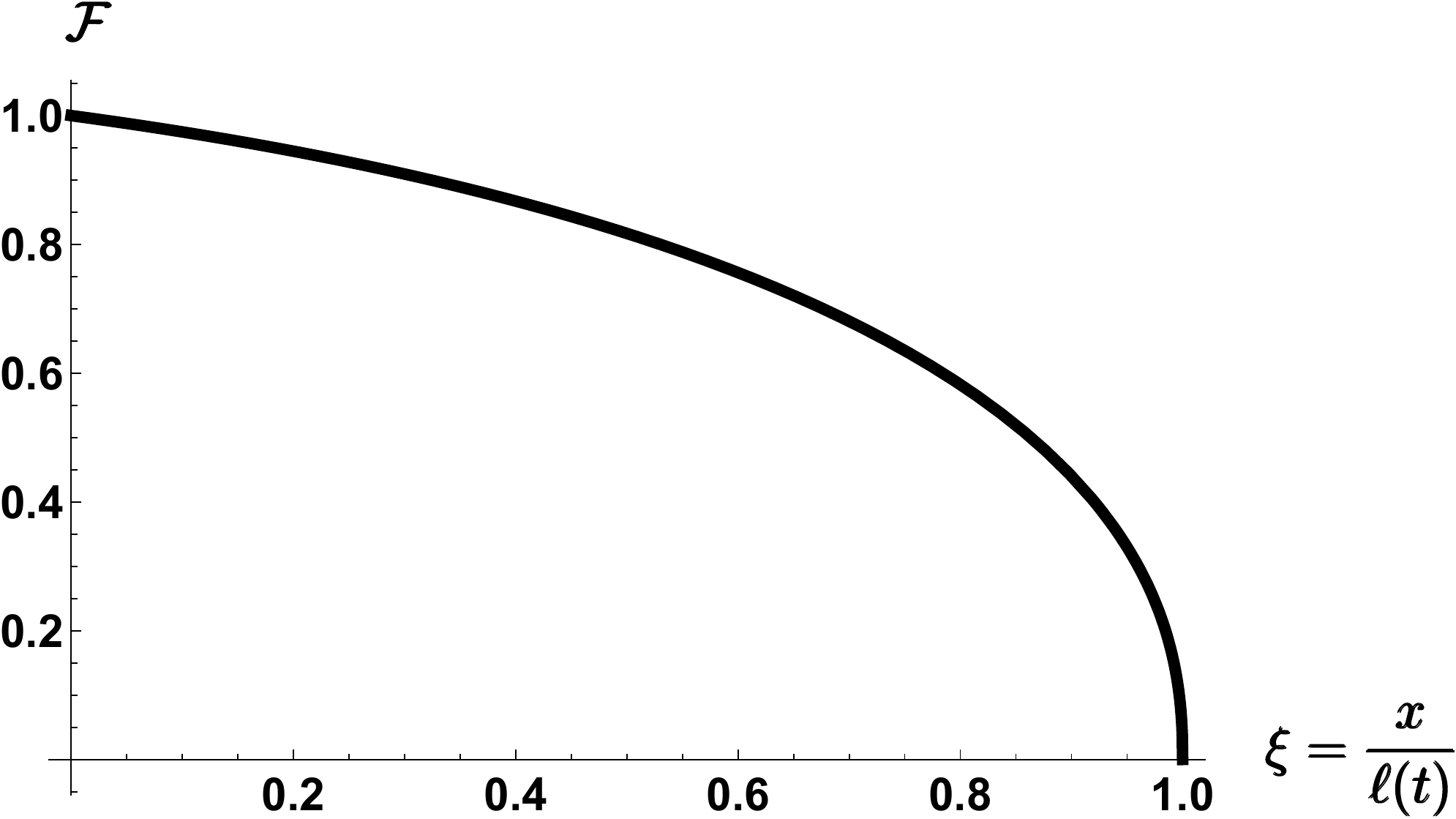}}
					\caption{Change of $\mathcal{F(\xi)}$ versus $\xi$. }
					\label{Figure 13}
				\end{figure}

				Finally, the Reynolds number discussed so far is determined by the fluid flow only in the neighborhood of the injection point. This Reynolds number, however, may not represent the flow near the fracture tip, where there is a switch from a Reynolds number dominated by the fracture depth $H$ to a local Reynolds number where the small width of the fracture dominates. As a result, there is a transition along the crack where the flow regime switches from turbulent flow to laminar flow. Thus, we would like to know the ratio of the length of the turbulent regime over the laminar region. Due to the fact that $q$ is independent of $H$ and only depends on time through $x/\ell(t)$, the quantity $H Re$ is a unique function of $x/\ell(t)$ for a given fluid. This relationship is shown in Fig.~\ref{Figure 11}.
				Comparing the values presented in Fig.~\ref{Figure 11} shows that the value of Reynolds number near the crack tip where the value of $x/\ell(t)$ approaches to 1 is close to zero and the behavior of the fluid in that region is thus laminar. Moreover, Fig.~\ref{Figure 11} indicates that there is a similarity between all the graphs; indeed the only thing that changes from one plot to the other is the value of the kinematic viscosity $\mu/\rho$. Combining, then, Eq.~\ref{ImplicitFORMall} and the definition of Reynolds numbers gives
				
				\begin{equation}\label{Reynolds_crack_lengthplayer}
				\begin{split}
				Re &=Re^*\mathcal{F(\xi)},\\
				Re^* &=\frac{\rho q_{in}}{\mu H},\\
				\mathcal{F(\xi)} &=\left(1-\xi\right)^{\frac{3}{7}}\left(1+0.05497\xi\right)^{\frac{5}{3}}\left(1+0.21005\xi\right)^{\frac{1}{2}}.
				\end{split}
				\end{equation}
				The change of $\mathcal{F(\xi)}$, which determines the variation of $Re$ for different values of $\xi$ is presented in Fig.~\ref{Figure 13}. In order to change the order of magnitude of $Re$ compared to $Re^*$, the value of $\mathcal{F(\xi)}$ should drop at least one order of magnitude which occurs for $\xi>0.9970$. Similarly, a two order of magnitude drop corresponds to $\xi>0.999986$.

				With the variation of $Re$ along the fracture in mind, consider the example of water as an injecting fluid ($\mu=0.001$ Pa$\cdot$s and $\rho=1000$ kg m$^{-3}$). The characteristic Reynolds number is $Re^*=10^6q_{in}/H$. If the depth is 20 m and the injecting fluid flow is $0.2$ m$^3$ s$^{-1}$, then $Re^*$ is $10^5$. Thus the fluid flow regime based on the characteristic Reynolds number is turbulent. However, as discussed, $Re$ decreases along the fracture, reaching a transition value of $Re\approx 2000$ at around $x/\ell(t)>0.98$. This indicates that about 98\% of the HF length is either in transition or turbulent regime. Thus, the GMS approximation is accurate enough for practical purposes in the transition regime. Due to the fact that HFs have a relatively large roughness scale compared to the fracture opening, in this case a valid global approximate solution can be obtained neglecting the laminar region near the tip. In other words, this example would correspond to a valid application of the present model. In contrast, if fluid flow is  $0.2$ m$^{3}$ s$^{-1}$ and the height of the crack is $70$ m, then $Re^*\approx 2900$, which is in transition to turbulent regime. In this case, Reynolds number drops to less than 2000 at about $x/\ell(t)>0.68$. In this case, then, 68\% of the HF is either in transition or is in turbulent regime which shows that around 1/3 of the crack is still in the laminar regime. In this latter example an approach considering the presence of both a turbulent and laminar region within the HF would be required.
				
				\section{Conclusions}
				
				The flow regime for some HF treatments is turbulent over the vast majority of the HF length. In particular, high-rate, water-driven HFs, as well as CO$_2$ driven fractures, tend to this regime. This is in contrast to the lower-rate, gel-driven fractures which comprised the main interest during the development of many HF solutions based on laminar flow models. Here we have presented a model for a blade-shaped (PKN) geometry HF growing in an impermeable rock and driven by a turbulent fluid. We derive a semi-analytical solution which: (a) embeds all rock, fluid, and geometric parameters in a scaling so that the resulting ode can be solved once for all cases, and (b) provides an accurate solution keeping only 2 terms of a polynomial series solution. The rapid convergence is enabled by embedding the near-tip behavior, also solved here in the course of the solution method, in the form of the polynomials. Failure to recognize the appropriate flow regime will lead to erroneous application of models based on laminar flow. Incorrect models are estimated to over predict the fracture length and under predict the fracture width and pressure by 40-50\%. As such, this model not only provides a benchmark solution for numerical simulation and a means for rapid estimation of fracture dimensions. It also provides impetus for ongoing research including experimental studies to find the most appropriate values of parameters $m$ and $\alpha$ for turbulent within a rough-walled deformable slot such as is encountered in HF applications.
				
				Here we show that using laminar flow instead of turbulent flow under conditions where most of the HF has Re>2500 can lead to enormous errors in calculating the fracture opening (i.e. more than 100\% at 1000 seconds of injection). And also a sizable error ($>$50\% as shown in Fig.~\ref{Figure 8}) is induced on the crack length and fluid pressure estimation. Ongoing efforts are aimed at expanding the ability of the model to consider the turbulent regime when it is appropriate to use a different form other than the generalized GMS equation.  It is expected this will be particularly necessary under conditions where proppant transport and the use of rheological-modifying additives are considered.
				\section{Acknowledgment}
				Navid zolfaghari and Andrew P. Bunger wish to acknowledge the support from the University of Pittsburgh Swanson School and Engineering and Center for Energy. Our discussions with Professor Jim Rice are also gratefully acknowledged.

				\pagebreak
				\begin{appendices}
					\renewcommand\thesection{}
					\section{Appendix I. Integrated Fluid Flow Equation}\label{q_simplified_app}
\begin{figure}[H]
						\centering
						\centerline{\includegraphics[height=40mm]{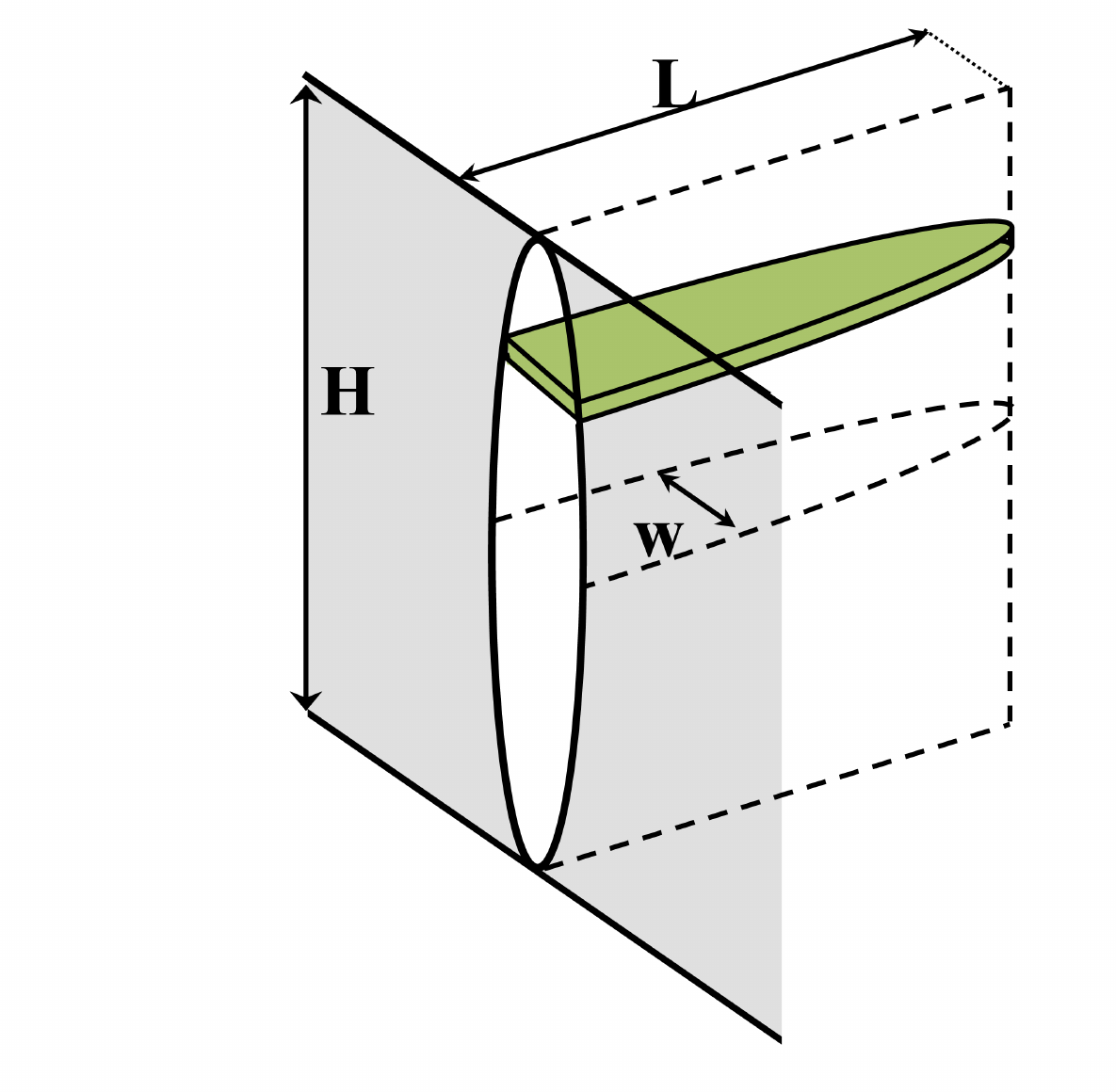}}
						\caption{Representation of two dimensional element in three dimensional crack. The integration of infinitesimal elements over the height of the crack will give the flow through a given cross section at location $x$.}
						\label{Figure 12}
					\end{figure}

					From elasticity and by assuming that the cross section of the crack is elliptical, it is possible to say that \cite{Nordgren72}
					\begin{equation}\label{elasticity_eq}
					W(x,z,t)=\frac{1-\nu}{G}(H^2-4z^2)^{1/2}(p-\sigma).
					\end{equation}
					At $z=0$ the maximum opening is given as
					\begin{equation}\label{elasticity_eq_z_zero}
					W(x,0,t)=\omega(x,t) =\frac{1-\nu}{G}H\left(p(x,t)-\sigma\right).
					\end{equation}
					Thus, the pressure gradient is
					\begin{equation}\label{elasticity_p}
					\frac{\partial p}{\partial x} =\frac{G}{1-\nu}\frac{1}{H}\frac{\partial \omega(x,t)}{\partial x}. 
					\end{equation}
					Inserting the crack opening expression Eq.~\ref{elasticity_eq} into the Gaukler-Manning-Strickler parametrization Eq.~\ref{General_Manning_strickler} gives
					\begin{equation}\label{f_p_comb_eq2}
					q_{2D}=\left[\frac{4}{\rho mk^{\alpha}}\omega^{3+\alpha}\left(1-\left(\frac{2z}{H}\right)^2\right)^{\frac{3+\alpha}{2}}\left(-\frac{\partial p}{\partial x}\right)\right]^{1/2},
					\end{equation}
					which is the two-dimensional flow at every height $z$, viz. Fig. \ref{Figure 12}. We can find the total flow rate by 
					integrating $q_{2D}$ over the height of the fracture, which gives
					\begin{equation}\label{int_three_dim}
					q=\int_{-H/2}^{H/2}q_{2D}dz=\left(\frac{4}{\rho mk^{\alpha}}\frac{G}{1-\nu}\frac{1}{H}\right)^{1/2}\left(-\omega^{3+\alpha}\frac{\partial\omega}{\partial x}\right)^{1/2}\int_{-H/2}^{H/2}\left[1-\left(\frac{2z}{H}\right)^2\right]^{\frac{(3+\alpha)}{4}}dz.
					\end{equation}
					Now letting $\varphi=(3+\alpha)/4$ and making the change of variable $\sin\theta=2z/H$ we have
					\begin{equation}\label{int_flow_four_dim_lambda}
					q=\left(\frac{4}{\rho mk^{\alpha}}\frac{G}{1-\nu}\frac{1}{H}\right)^{1/2}\omega^{2\varphi}\left(-\frac{\partial\omega}{\partial x}\right)^{1/2}\frac{H}{2}\int_{-\pi/2}^{\pi/2}(\cos{\theta})^{2\varphi+1} d\theta.
					\end{equation}
					Thus, we can write the total flux $q$ in terms of the Beta function $\beta(x,y)$ as
					\begin{equation}\label{int_flow_five_dim_lambda}
					q=\frac{H}{2}\left(\frac{4}{\rho mk^{\alpha}}\frac{G}{1-\nu}\frac{1}{H}\right)^{1/2}\omega^{2\varphi}\left(-\frac{\partial \omega}{\partial x}\right)^{1/2}\beta\left(\frac{1}{2},\varphi+1\right),
					\end{equation}
					where the Beta function is defined as \cite{Abbramovitz64}
					\begin{equation}\label{beta_func_orig}
					\beta(x,y)=\int_0^{1}{t^{x-1}(1-t)^{y-1}}dt,
					\end{equation}
					or
					\begin{equation}
					\beta(x,y)=\frac{\Gamma(x)\Gamma(y)}{\Gamma(x+y)},
					\end{equation}
					for $\mathbb{R}(x)>0$ and $\mathbb{R}(y)>0$ and $\Gamma(x)$ is the Gamma function~\cite{Abbramovitz64}.
					\section{Appendix II. Supplement Data}
					\begin{enumerate}
						\item Turbulent flow.
						\item Laminar to turbulent transition model.
						\item Near tip full asymptotics.
						\item Alternative scaling approach.
					\end{enumerate}
					\section{Appendix III. Notation}
					\label{Nomencluture}
					\emph{The following symbols are used in this paper:}
					\nopagebreak
					\par
\begin{tabu}{r  @{\hspace{1em}=\hspace{1em}}  l}
		   	$A$ & $\text{Cross-sectional area of the crack at coordinate x and time t;}$\\
		   	$B$ & $\text{Beta Function;}$\\
		   	$f_{p}$ & $\text{Darcy-Weisbach friction factor;}$\\
		   	$G$ & $\text{Shear modulus of elasticity of the rock;}$\\ 
		   	$G_i(p,q,r)$ & $i^{th}\text{~order Jacobi polynomial;}$\\
		   	$H$ & $\text{Height of the stimulating rock;}$\\
		   	$h_i(p,q)$ & $\text{Norm of $i^{th}$ order Jacobi polynomial;}$\\
		   	$k$ & $\text{Nikuradse channel wall roughness}$ \\
		   	$\ell(t)$ & $\text{Length of the crack at time t;}$\\
		   	$\ell_N(t)$ & $\text{Length of the crack calculated based on Nordgren solution;}$\\
		   	$m$ & $\text{Constant coefficient for modelling turbulent flow;}$\\
		   	$n$ & $\text{Number of terms to truncate Jacobian polynomial series;}$\\
		   	$p$ & $\text{Fluid pressure;}$\\
		   	$p_{net}$ & $\text{Net pressure inside the crack and is equal to $p-\sigma$} $\\
		   	$p_{net_N}$ & $\text{Net pressure calculated based on Nordgren solution;}$\\
		   	$Q$ & $\text{Scaled Flux;}$\\
		   	$q$ & $\text{Fluid flow within the fracture from a cross-section at location x;}$\\
		   	$q_{in}$ & $\text{Volume rate of flow at inlet;}$\\
		   	$q_{2D}$ & $\text{Unidirectional fluid flow through the two dimensional channel;}$\\
		   	$R(x)$ & $\text{The weight function in general definition of orthogonal polynomials;}$\\
		   	$R_h$ & $\text{Hydraulic radius;}$\\
		   	$Re$ & $\text{Reynold number at location x and at time t;}$\\
		   	$Re^*$ & $\text{Characteristic Reynold number;}$\\
		   	$t$ & $\text{Time;}$\\
		   	$W$ & $\text{Opening of PKN crack at location (x,z) and at time t;}$\\
		   	$(x,z)$ & $\text{Coordinate value at x-axis and z-axis;}$\\
		   	$\alpha$ & $\text{Constant coefficient that appear as power for modelling turbulent flow;}$\\
		   	$\gamma(t)$ & $\text{Dimensionless length of the crack;}$\\
		   	$\Gamma$ & $\text{Gamma function;}$\\
		   	$\delta_{ij}$ & $\text{Kronecker delta;}$\\
		   	$\xi$ & $\text{Scaled coordinate;}$\\
		   	$\Lambda$ & $\text{Pre-defined constant that is mentioned in Eq.~\ref{q_simplified};}$\\
		   	$\mu$ & $\text{Fluid viscosity;}$\\
		   	$\nu$ & $\text{Poisson's ratio of the rock;}$\\
		   	$\rho$ & $\text{Fluid density;}$\\              
		   	$\Xi$ & $\text{Pre-defined constant ($\Xi=\frac{4\Lambda\Upsilon}{\pi H}$);}$\\
		   	$\varrho$ & $\text{Constant depending on $\alpha$ and can be find from Eq.~\ref{Figure 1};}$\\
		   	$\sigma$ & $\text{In-situ stress;}$\\
		   		$\Upsilon$ & $\text{pre-defined constant that is mentioned in Eq.~\ref{q_simplified};}$\\
		   		$\varphi$ & $\text{Constant depending on $\alpha$ and can be find from Eq.~\ref{q_simplified};}$\\
		   		$\omega$ & $\text{Maximum opening in each vertical cross-section at time t;}$\\
		   		$\overline{\omega}$ & $\text{Average value of opening at each vertical cross-section at time t;}$\\
		   		$\omega_N$ & $\text{Maximum opening calculated based on Nordgren solution;}$\\
		   			$\Omega(\xi)$ & $\text{Dimensionless opening;}$\\
		   			$\hat{\Omega}_i$ & $\text{Base function to define dimensionless opening;}$\\
	\end{tabu} \\

\center\indent\begin{tabu}{r  @{\hspace{1em}=\hspace{1em}}  l}
		$\mathcal{A}_i$ & $\text{Unknown coefficients to define dimensionless opening;}$\\
		$\mathcal{B}$ & $\text{Constant that can be calculated from Eq.~\ref{int_omega_solved};}$\\
		$\mathcal{C}_1$ & $\text{Constant that is dependant on $\alpha$ and can be obtained from Fig.~\ref{Figure 5};}$\\
		$\mathcal{C}_2$ & $\text{Constant that is dependant on $\alpha$ and can be obtained from Fig.~\ref{Figure 5};}$\\
		$\mathfrak{D}_i$ & $\text{Proper constant to be chosen to satisfy orthogonality;}$\\
		$\mathfrak{f}_i$ & $\text{Proper set of function to satisfy Eq.~\ref{Orthogonalitytwo};}$\\
		$\mathcal{F}(\xi)$ & $\text{A function that defines the decay of Re along the crack (see Eq.\ref{Reynolds_crack_lengthplayer});}$\\
		$\mathfrak{P}$ & $\text{Wetted perimeter of the crack cross section;}$\\
		$\mathcal{W}(t)$ & $\text{Characteristic width of HF;}$\\
		$\mathcal{X}$ & $\text{Constant to define crack tip (see Eq.~\ref{int_omega_solved});}$\\
	\end{tabu} 

			\end{appendices}	
			\newpage
		\bibliography{JFM2015_NA_v2}

\begin{thebibliography}{}

\bibitem[\protect\citeauthoryear{}{Abramowitz and Stegun}{1972}]{Abbramovitz64}
Abramowitz, M. and Stegun, I.~A. (1972).
\newblock {\em Handbook of Mathematical Functions: with Formulas, Graphs, and
  Mathematical Tables}, Vol.~10.
\newblock Dover, New York.

\bibitem[\protect\citeauthoryear{}{Adachi and
  Detournay}{2002}]{adachidetornay2002}
Adachi, J.~I. and Detournay, E. (2002).
\newblock ``Self-similar solution of a plane-strain fracture driven by a
  power-law fluid.''\ {\em International Journal for Numerical and Analytical
  Methods in Geomechanics}, 26(6), 579--604.

\bibitem[\protect\citeauthoryear{}{Adachi and
  Peirce}{2008}]{adachi2008asymptotic}
Adachi, J.~I. and Peirce, A.~P. (2008).
\newblock ``Asymptotic analysis of an elasticity equation for a finger-like
  hydraulic fracture.''\ {\em Journal of Elasticity}, 90(1), 43--69.

\bibitem[\protect\citeauthoryear{}{Ames and Bunger}{2015}]{AmBu15}
Ames, B.~C. and Bunger, A.~P. (2015).
\newblock ``Role of turbulent flow in generating short hydraulic fractures with
  high net pressure in slickwater treatments.''\ {\em SPE Hydraulic Fracturing
  Technology Conference}, number SPE 173373, Society of Petroleum Engineers.

\bibitem[\protect\citeauthoryear{}{Anthonyrajah
  et~al.\@}{2013}]{Anthonyrajah2013}
Anthonyrajah, M., Mason, D.~P., and Fareo, A.~G. (2013).
\newblock ``Propagation of a pre-existing turbulent fluid fracture.''\ {\em
  International Journal of Non-Linear Mechanics}, 54, 105--114.

\bibitem[\protect\citeauthoryear{}{Barenblatt}{1996}]{Barenblatt1996}
Barenblatt, G.~I. (1996).
\newblock {\em Scaling, self-similarity, and intermediate asymptotics:
  dimensional analysis and intermediate asymptotics}, Vol.~14.
\newblock Cambridge University Press.

\bibitem[\protect\citeauthoryear{}{Bunger and
  Detournay}{2007}]{bungerdetornay2007}
Bunger, A.~P. and Detournay, E. (2007).
\newblock ``Early-time solution for a radial hydraulic fracture.''\ {\em
  Journal of engineering mechanics}, 133(5), 534--540.

\bibitem[\protect\citeauthoryear{}{Darcy}{1857}]{Darcy1857}
Darcy, H. (1857).
\newblock {\em Recherches exp{\'e}rimentales relatives au mouvement de l'eau
  dans les tuyaux}, Vol.~1.
\newblock Mallet-Bachelier.

\bibitem[\protect\citeauthoryear{}{De~Pater}{2015}]{de2015s}
De~Pater, H.~J. (2015).
\newblock ``Hydraulic fracture containment: New insights into mapped
  geometry.''\ {\em SPE Hydraulic Fracturing Technology Conference}, Society of
  Petroleum Engineers.

\bibitem[\protect\citeauthoryear{}{Dontsov}{2016}]{Dontsov2016}
Dontsov, E.~V. (2016).
\newblock ``Tip region of a hydraulic fracture driven by a laminar-to-turbulent
  fluid flow.''\ {\em Journal of Fluid Mechanics}, 797, R2 (12 pages).

\bibitem[\protect\citeauthoryear{}{Emanuele et~al.\@}{1998}]{emanuele1998}
Emanuele, M., Minner, W., Weijers, L., Broussard, E., Blevens, D., Taylor, B.,
  et~al.\@ (1998).
\newblock ``A case history: completion and stimulation of horizontal wells with
  multiple transverse hydraulic fractures in the lost hills diatomite.''\ {\em
  SPE Rocky Mountain Regional/Low-Permeability Reservoirs Symposium}, Society
  of Petroleum Engineers.

\bibitem[\protect\citeauthoryear{}{Emerman et~al.\@}{1986}]{Emerman1986}
Emerman, S.~H., Turcotte, D., and Spence, D. (1986).
\newblock ``Transport of magma and hydrothermal solutions by laminar and
  turbulent fluid fracture.''\ {\em Physics of the earth and planetary
  interiors}, 41(4), 249--259.

\bibitem[\protect\citeauthoryear{}{Emmons}{1951}]{emmons2012}
Emmons, H.~W. (1951).
\newblock ``The laminar-turbulent transition in a boundary layer-part {I}.''\
  {\em Journal of the Aeronautical Sciences}, 18(7), 490--498.

\bibitem[\protect\citeauthoryear{}{Fischer et~al.\@}{2008}]{fischer2008}
Fischer, T., Hainzl, S., Eisner, L., Shapiro, S., and Le~Calvez, J. (2008).
\newblock ``Microseismic signatures of hydraulic fracture growth in sediment
  formations: Observations and modeling.''\ {\em Journal of Geophysical
  Research: Solid Earth}, 113(B2).

\bibitem[\protect\citeauthoryear{}{Fisher and
  Warpinski}{2012}]{fisherwarpinski2012}
Fisher, M.~K. and Warpinski, N.~R. (2012).
\newblock ``Hydraulic-fracture-height growth: Real data.''\ {\em SPE Production
  \& Operations}, 27(1), 8--19.

\bibitem[\protect\citeauthoryear{}{Gauckler}{1867}]{gauckler}
Gauckler, P. (1867).
\newblock ``Etudes th{\'e}oriques et pratiques sur l'ecoulement et le mouvement
  des eaux.''\ 64, 818--822.

\bibitem[\protect\citeauthoryear{}{Geertsma and
  De~Klerk}{1969}]{geertsma1969rapid}
Geertsma, J. and De~Klerk, F. (1969).
\newblock ``A rapid method of predicting width and extent of hydraulically
  induced fractures.''\ {\em Journal of Petroleum Technology}, 21(12), 1--571.

\bibitem[\protect\citeauthoryear{}{Gioia and Chakraborty}{2006}]{gioia2006}
Gioia, G. and Chakraborty, P. (2006).
\newblock ``Turbulent friction in rough pipes and the energy spectrum of the
  phenomenological theory.''\ {\em Physical review letters}, 96(4), 044502.

\bibitem[\protect\citeauthoryear{}{Henderson}{1966}]{henderson1966open}
Henderson, F.~M. (1966).
\newblock {\em Open channel flow}.
\newblock Macmillan series in civil engineering. Macmillan.

\bibitem[\protect\citeauthoryear{}{Huppert}{1982}]{Huppert1982}
Huppert, H.~E. (1982).
\newblock ``The propagation of two-dimensional and axisymmetric viscous gravity
  currents over a rigid horizontal surface.''\ {\em Journal of Fluid
  Mechanics}, 121, 43--58.

\bibitem[\protect\citeauthoryear{}{Kano et~al.\@}{2015}]{Kano15}
Kano, M., Zolfaghari, N., Ames, B.~C., and Bunger, A.~P. (2015).
\newblock ``Solution for a pkn hydraulic fracture driven by turbulent fluid
  with large leakoff.''\ {\em Hydraulic Fracturing Journal}, 2(1), 34--38.

\bibitem[\protect\citeauthoryear{}{King}{2010}]{king2010}
King, G.~E. (2010).
\newblock ``Thirty years of gas shale fracturing: what have we learned?.''\
  {\em SPE Annual Technical Conference and Exhibition}, Society of Petroleum
  Engineers.

\bibitem[\protect\citeauthoryear{}{Lister}{1990}]{Lister1990}
Lister, J.~R. (1990).
\newblock ``Buoyancy-driven fluid fracture: similarity solutions for the
  horizontal and vertical propagation of fluid-filled cracks.''\ {\em Journal
  of Fluid Mechanics}, 217, 213--239.

\bibitem[\protect\citeauthoryear{}{Lister and Kerr}{1991}]{ListerKerr1991}
Lister, J.~R. and Kerr, R.~C. (1991).
\newblock ``Fluid-mechanical models of crack propagation and their application
  to magma transport in dykes.''\ {\em Journal of Geophysical Research: Solid
  Earth}, 96(B6), 10049--10077.

\bibitem[\protect\citeauthoryear{}{Manning}{1891}]{Manning1891}
Manning, R. (1891).
\newblock ``On the flow of water in open channels and pipes.''\ {\em
  Transactions of the Institution of Civil Engineers of Ireland}, 20, 161--207.

\bibitem[\protect\citeauthoryear{}{Maxwell et~al.\@}{2002}]{maxwell2002}
Maxwell, S.~C., Urbancic, T., Steinsberger, N., Zinno, R., et~al.\@ (2002).
\newblock ``Microseismic imaging of hydraulic fracture complexity in the
  barnett shale.''\ {\em SPE annual technical conference and exhibition},
  Society of Petroleum Engineers.

\bibitem[\protect\citeauthoryear{}{Mayerhofer et~al.\@}{2000}]{mayerhofer2000}
Mayerhofer, M.~J., Walker~Jr, R.~N., Urbancic, T., Rutledge, J.~T., et~al.\@
  (2000).
\newblock ``East texas hydraulic fracture imaging project: Measuring hydraulic
  fracture growth of conventional sandfracs and waterfracs.''\ {\em SPE Annual
  Technical Conference and Exhibition}, Society of Petroleum Engineers.

\bibitem[\protect\citeauthoryear{}{Munson et~al.\@}{2002}]{munson02}
Munson, B.~R., Young, B.~F., and Okiishi, T.~H. (2002).
\newblock {\em Fundamentals of Fluid Mechanics.}
\newblock John Wiley and Sons, United States of America, 4 edition.

\bibitem[\protect\citeauthoryear{}{Murdoch and Slack}{2002}]{murdoch2002}
Murdoch, L.~C. and Slack, W.~W. (2002).
\newblock ``Forms of hydraulic fractures in shallow fine-grained formations.''\
  {\em Journal of Geotechnical and Geoenvironmental Engineering}, 128(6),
  479--487.

\bibitem[\protect\citeauthoryear{}{Nilson}{1981}]{Nilson1981gas}
Nilson, R. (1981).
\newblock ``Gas-driven fracture propagation.''\ {\em Journal of Applied
  Mechanics}, 48(4), 757--762.

\bibitem[\protect\citeauthoryear{}{Nilson}{1988}]{Nilson1988gas}
Nilson, R. (1988).
\newblock ``Similarity solutions for wedge-shaped hydraulic fractures driven
  into a permeable medium by a constant inlet pressure.''\ {\em International
  Journal for Numerical and Analytical Methods in Geomechanics}, 12(5),
  477--495.

\bibitem[\protect\citeauthoryear{}{Nordgren}{1972}]{Nordgren72}
Nordgren, R. (1972).
\newblock ``Propagation of a vertical hydraulic fracture.''\ {\em Society of
  Petroleum Engineers Journal}, 12(4), 306--314.

\bibitem[\protect\citeauthoryear{}{Perkins and Kern}{1961}]{perkins61}
Perkins, T.~K. and Kern, L.~R. (1961).
\newblock ``Widths of hydraulic fractures.''\ {\em Journal of Petroleum
  Technology}, 13(9), 937--949.

\bibitem[\protect\citeauthoryear{}{Rice et~al.\@}{2015}]{tsai2015model}
Rice, J.~R., Tsai, V.~C., Fernandes, M.~C., and Platt, J.~D. (2015).
\newblock ``Time scale for rapid draining of a surficial lake into the
  greenland ice sheet.''\ {\em Journal of Applied Mechanics}, 82(7), 071001.

\bibitem[\protect\citeauthoryear{}{Rutledge and Phillips}{2003}]{rutledge2003}
Rutledge, J.~T. and Phillips, W.~S. (2003).
\newblock ``Hydraulic stimulation of natural fractures as revealed by induced
  microearthquakes, carthage cotton valley gas field, east texas.''\ {\em
  Geophysics}, 68(2), 441--452.

\bibitem[\protect\citeauthoryear{}{Savitski and Detournay}{2002}]{savitski2002}
Savitski, A.~A. and Detournay, E. (2002).
\newblock ``Propagation of a penny-shaped fluid-driven fracture in an
  impermeable rock: asymptotic solutions.''\ {\em International Journal of
  Solids and Structures}, 39(26), 6311--6337.

\bibitem[\protect\citeauthoryear{}{Strickler}{1923}]{strickler23}
Strickler, A. (1923).
\newblock {\em Beitr{\"a}ge zur Frage der Geschwindigkeitsformel und der
  Rauhigkeitszahlen f{\"u}r Str{\"o}me, Kan{\"a}le und geschlossene Leitungen}.
\newblock Technical Report Tech. Rep. 16, Mitteilungen des Eidgenossischen
  Amtes f{\"u}r Wasserwirtschaft, Bern.

\bibitem[\protect\citeauthoryear{}{Te~Chow}{1959}]{te1959open}
Te~Chow, V. (1959).
\newblock {\em Open channel hydraulics}.
\newblock McGraw-Hill Book Company, Inc; New York.

\bibitem[\protect\citeauthoryear{}{Tsai and Rice}{2010}]{tsai2010model}
Tsai, V.~C. and Rice, J.~R. (2010).
\newblock ``A model for turbulent hydraulic fracture and application to crack
  propagation at glacier beds.''\ {\em Journal of Geophysical Research: Earth
  Surface}, 115(F3).

\bibitem[\protect\citeauthoryear{}{Tsai and Rice}{2012}]{tsai2012model}
Tsai, V.~C. and Rice, J.~R. (2012).
\newblock ``Modeling turbulent hydraulic fracture near a free surface.''\ {\em
  Journal of Applied Mechanics}, 79(3), 031003.

\bibitem[\protect\citeauthoryear{}{Warpinski et~al.\@}{1999}]{warpinski1999}
Warpinski, N., Branagan, P., Mahrer, K., Wolhart, S., and Moschovidis, Z.
  (1999).
\newblock ``Microseismic monitoring of the mounds drill cuttings injection
  tests.''\ {\em Proc},  1025--1032.

\bibitem[\protect\citeauthoryear{}{Weisbach}{1855}]{weisbach1855}
Weisbach, J.~L. (1855).
\newblock {\em Die Experimental Hydraulik}.
\newblock Engelhardt.

\bibitem[\protect\citeauthoryear{}{Wright et~al.\@}{1999}]{wright1999}
Wright, C., Weijers, L., Davis, E., and Mayerhofer, M. (1999).
\newblock ``Understanding hydraulic fracture growth: tricky but not
  hopeless.''\ {\em SPE annual technical conference},  661--670.

\bibitem[\protect\citeauthoryear{}{Zia and Lecampion}{2016}]{zia2016laminar}
Zia, H. and Lecampion, B. (2016).
\newblock ``Laminar-turbulent transition in the propagation of height-contained
  hydraulic fracture.

\end{thebibliography}
		
		\pagebreak
		\setcounter{equation}{0}
		\setcounter{figure}{0}
		\setcounter{table}{0}
		\setcounter{page}{1}
		\setcounter{section}{0}
		\makeatletter
		\renewcommand{\theequation}{S\arabic{equation}}
		\renewcommand{\thefigure}{S\arabic{figure}}
		\renewcommand{\thesection}{S\arabic{section}}
		\title{Supplement to: Blade-shaped (PKN) Hydraulic Fracture Driven By A Turbulent Fluid In An Impermeable Rock}	
		%
		%
		\maketitle
		\section{Turbulent Flow}\label{turb}
\begin{figure}[H]
			\centering
			\includegraphics[angle=0,width=140mm]{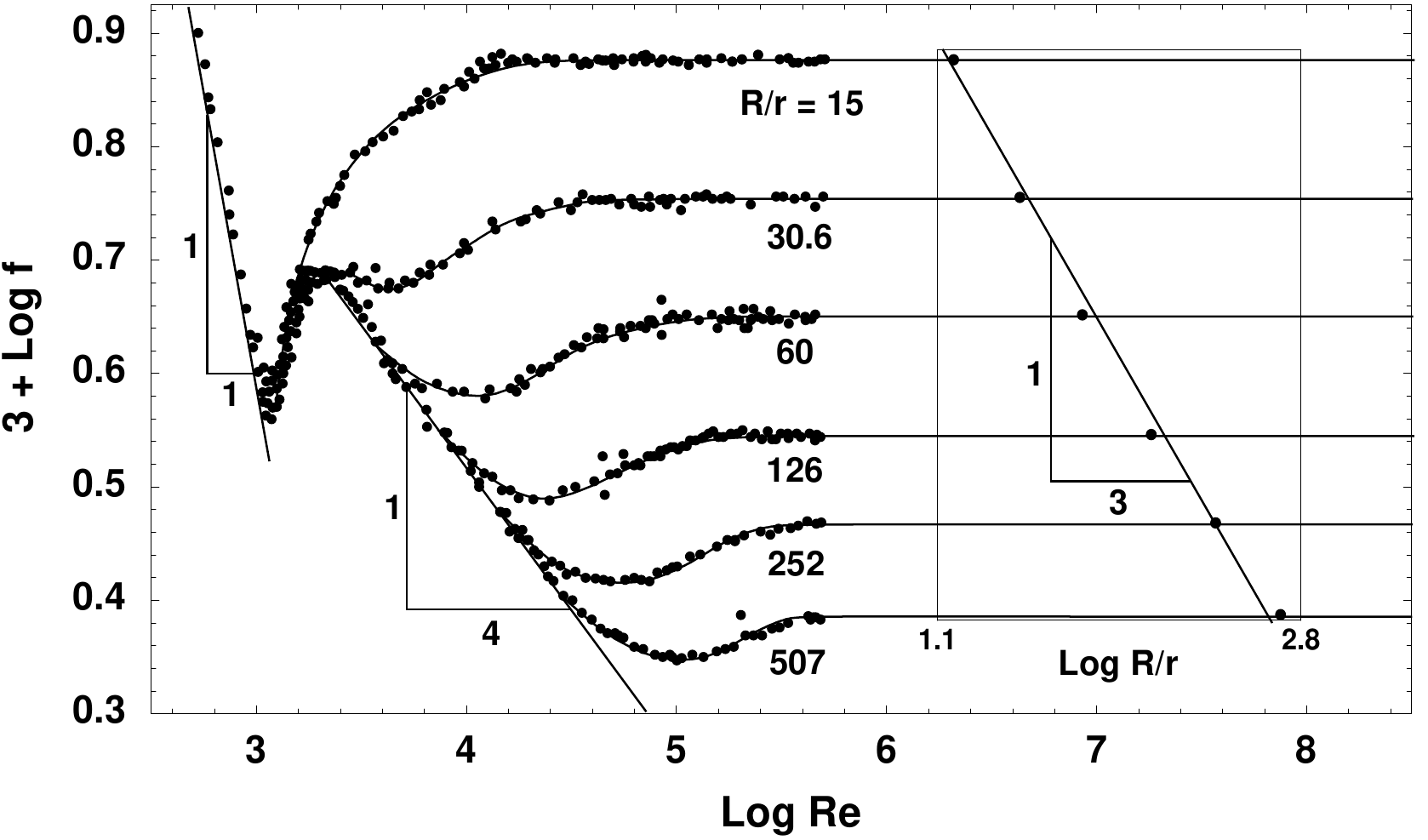}
			\caption{Moody diagram for pipe flow~\protect\cite{gioia2006}, with permission.}
			\label{moody}
		\end{figure}

		A better understanding of the physics of turbulent flows would allow for the determination of the pressure drop in a turbulent flow through a crack. This section uses dimensional analysis to make an analogy between flow through a crack and flow through a pipe. Most studies of HF with turbulent flow have only been carried out with either pipe flow assumption or channel flow, which fails to resolve the contradiction that crack growth cannot be categorized as either pipe flow, nor channel flow assumptions. This study is unable to encompass the entire discussion about turbulent flow; however, using dimensional analysis to obtain flow equation for turbulent regime can enhance the clarity of this method.  
		The fluid pressure drop inside the pipe depends on different parameters and can be described as $\Delta P=\mathcal{F}(V,D,l,k,\mu,\rho)$ where $V$ is the mean velocity of the fluid, $D$ is the pipe diameter, $l$ is the pipe length, $k$ is the pipe roughness, $\mu$ is the fluid viscosity, and $\rho$ is the fluid density. The Buckingham-$\Pi$ theorem gives that
		\begin{equation}\label{pi_non_dim}
		\frac{\partial P}{\partial x}=\rho V^2 \frac{1}{D}\mathcal{F}(Re,\frac{k}{D}),
		\end{equation}
		by replacing the mean velocity $V$ with $q/D$ and re-arranging the equation, we find that 
		\begin{equation}\label{pi_non_dim_two}
		q=\left(-\frac{D^3}{\rho\mathcal{F}}\frac{\partial P}{\partial x}\right)^{1/2}, 
		\end{equation} 
		where $\mathcal{F}$ depends on Reynolds number and roughness of the crack. This equation is the same as for GMS (eqn 4 in the paper) where $\mathcal{F}$ is assumed to be independent of Reynolds number and a power law in roughness given by:
		\begin{equation}\label{WeisbachS}
		f_p=m\left(\frac{k}{W}\right)^{\alpha}.
		\end{equation} 
		This is clearly shown in the Moody diagram, figure \ref{moody}. For Reynolds numbers greater than about 10$^{5}$, the friction factor is nearly independent of Reynolds number. Additional roughness, however, affects the friction considerably and the inset to figure \ref{moody} shows that the friction varies as a power law in roughness.
		
		No-slip condition on the contact boundary of fluid with solid will cause a viscous layer to form. In this viscous layer, the shear force is mainly described based on viscosity of the fluid. On the other hand, the thickness of viscous layer has inverse relation to the value of Reynolds number. When the Reynolds number is very large, the viscous sublayer is very thin and the shear stress is mostly defined by Reynolds stress. In such a cases the shear stress is more in the format of $K\rho V^2/2$ where $K$ is a parameter that depends on roughness of the wall (see \citeNP{munson02} for more detail). Therefor, by increasing the Reynolds number, in contrast to laminar flow, the shear force is more dependent on fluid density and crack roughness rather than fluid viscosity. And that is why in turbulent equations like in Eq.~\ref{pi_non_dim_two} and \ref{WeisbachS}, the fluid density will appear but in Poiseuille equation, the viscosity will show up. This phenomena is also explainable based on the Moody diagram.
		In general, the Moody diagram is mainly developed to calculate the friction factor in the circular pipes. However, by introducing hydraulic radius, it is possible to connect non-circular cross sections and channels to the friction factor. For the laminar part of the Moody diagram, there is one unique line that explains the friction factor which confirms the fact that friction factor is only function of Reynolds number. However, by increasing Reynolds number, as was pointed out above, the viscous sublayer will become smaller. As this trend continuous, the height of the viscous sublayer will reduce until the length of the surface roughness is comparable with the height of the sublayer. Therefore, the friction factor will become function of two parameters: surface roughness and Reynolds number. In fact, this two parameters are acting as a counterbalance against each other. As an illustration, if Reynolds increases, the viscous sublayer will become smaller and friction factor reduces. On the other hand, the effect of surface roughness will become more important which will cause the friction factor to increase. Hence, the final value of friction factor is the combination of the effect of this two values.  This phenomena is illustrated in the Moody digram very well. After the laminar region, since the friction factor is function of two variables, it will not be possible to explain it with one plot and that is why there is several branches for different values of surface roughness. Besides, in Moody diagram, as Reynolds increase, the friction factor will decrease. And as surface roughness, increase, the friction factor will increase. So the increase of friction factor is due to either decreasing Reynolds number or increasing surface roughness.  
		
		For very large Reynolds numbers, the value of laminar sublayer diminish so much that the fracture toughness will become dominant and the effect of Reynolds number is eliminated. In such a cases, the friction factor will be only function of surface roughness $f(\varepsilon/D)$, in contrast to laminar flow which was only function of Reynolds number $f(Re)$. When the flow is just function of surface roughness, the fluid flow is called wholly turbulent flow. Looking at Moody diagram for turbulent flow, you can observe that first of all, as much as the surface roughness increases, the plot will become more flat, which means that for large span of Reynolds numbers and fixed value of surface roughness, the changes of friction factor will become smaller. That is, friction factor will become independent to Reynolds number and become more linked to surface roughness which indicate that wholly turbulent flow is happening. Also, for larger Reynolds numbers, the plot will again become flat that promises the happening of wholly turbulent region. 
		
		In hydraulic fracture, usually the surface roughness is big when compared to fracture opening. For example, the ratio of roughness height to fracture opening is in more than 0.1 ($\varepsilon/W>0.1$). So according to Moody diagram, the friction factor plot is almost horizontal line and is mostly in wholly turbulent region. And that is why mainly in GMS equation, the friction factor is only function of scaled roughness rather than Reynolds number.
		
		\section{Laminar to turbulent transition model}
		\subsection{Regime transition for different fluids based on general open channel fluid mechanics}\label{App diff fluids}
		We define a characteristic Reynolds number that depends on parameters such as the fluid density $\rho$, viscosity $\mu$, pumping rate  $q_{in}$, and the layer thickness $H$, which combine into
		\[Re=\frac{\rho q_{in}}{H\mu}.\]
		The fluid flow is assumed to be laminar for Reynolds number less than 500 and fully-developed turbulence for Reynolds numbers greater than 12500. From 500 to 12500 is the transition between laminar to turbulent flow. Typical heights of HFs fall in the range 20 m $<H<$ 200 m \cite{fisherwarpinski2012} and $\rho$, $\mu$ are material properties of the working fluid. We will consider injection rates between 0.01 m$^3$ s$^{-1}$ $<q_{in}<$ 0.2 m$^3$ s$^{-1}$. Thus, the ratio of injection rate to layer thickness falls in the range $5\times 10^{-5}$ m$^3$ s$^{-1}$ $<q_{in}/H<$ 0.01 m$^3$ s$^{-1}$.
		
\begin{figure}[H]
			\centering
			\includegraphics[height=140mm]{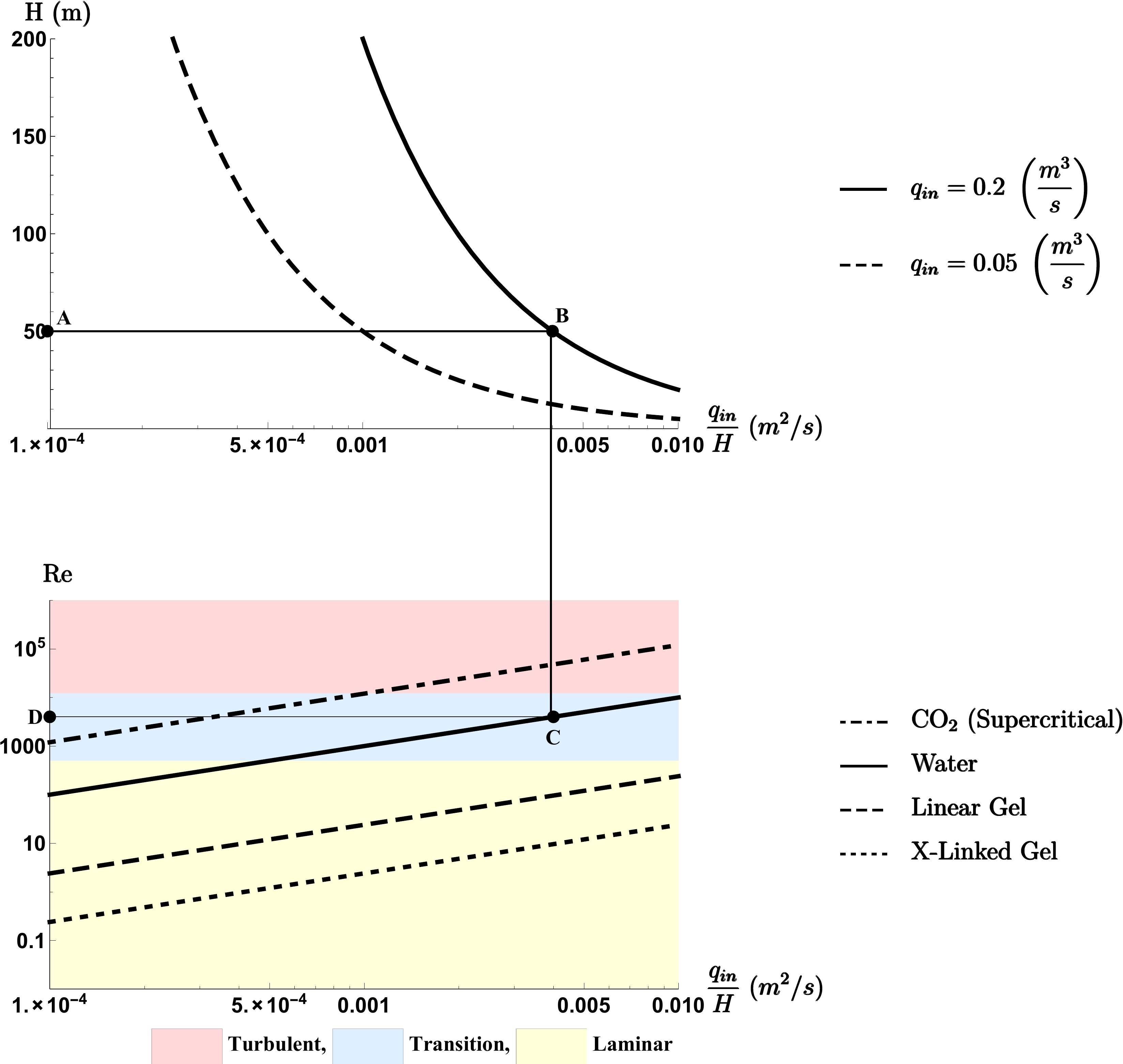}
			\caption{Variation of fluid flow regime by changing the fluid properties.}
			\label{Figure 9}
		\end{figure}

		The summary of the change of Reynolds number for different fluids is shown in Fig.~\ref{Figure 9}. For example, we can compute the Reynolds number for water pumped at $q_{in}=0.2$ m$^3$ s$^{-1}$ into a layer with height $H=50$ m. First, we mark point A at $H=50$, then we draw a horizontal line from point A until it hits the curve for $q_{in}=0.2$ m$^3$ s$^{-1}$, which we call point B. We then draw a vertical line from point B until it hits the line for water and we call it point C. Finally, we can draw a horizontal line from point C to the left y-axis to read the value of $\log(Re)$ (point D). The background color at point C indicates the regime and practical fluxes fall in the range $q_{in}=0.05$ m$^3$ s$^{-1}$ to $q_{in}=0.2$ m$^3$ s$^{-1}$. Thus, for all practical applications the flow regime for cross-linked and linear gel is laminar. Water, however, falls in the transition zone and CO$_2$ is transitional or fully-developed turbulence depending on $q_{in}/H$.
		\pagebreak
		\subsection{Regime transition for different fluids based on general open channel fluid mechanics, pressure approach}\label{App diff fluids new method}
\begin{figure}[H]
			\centering
			\includegraphics[height=140mm]{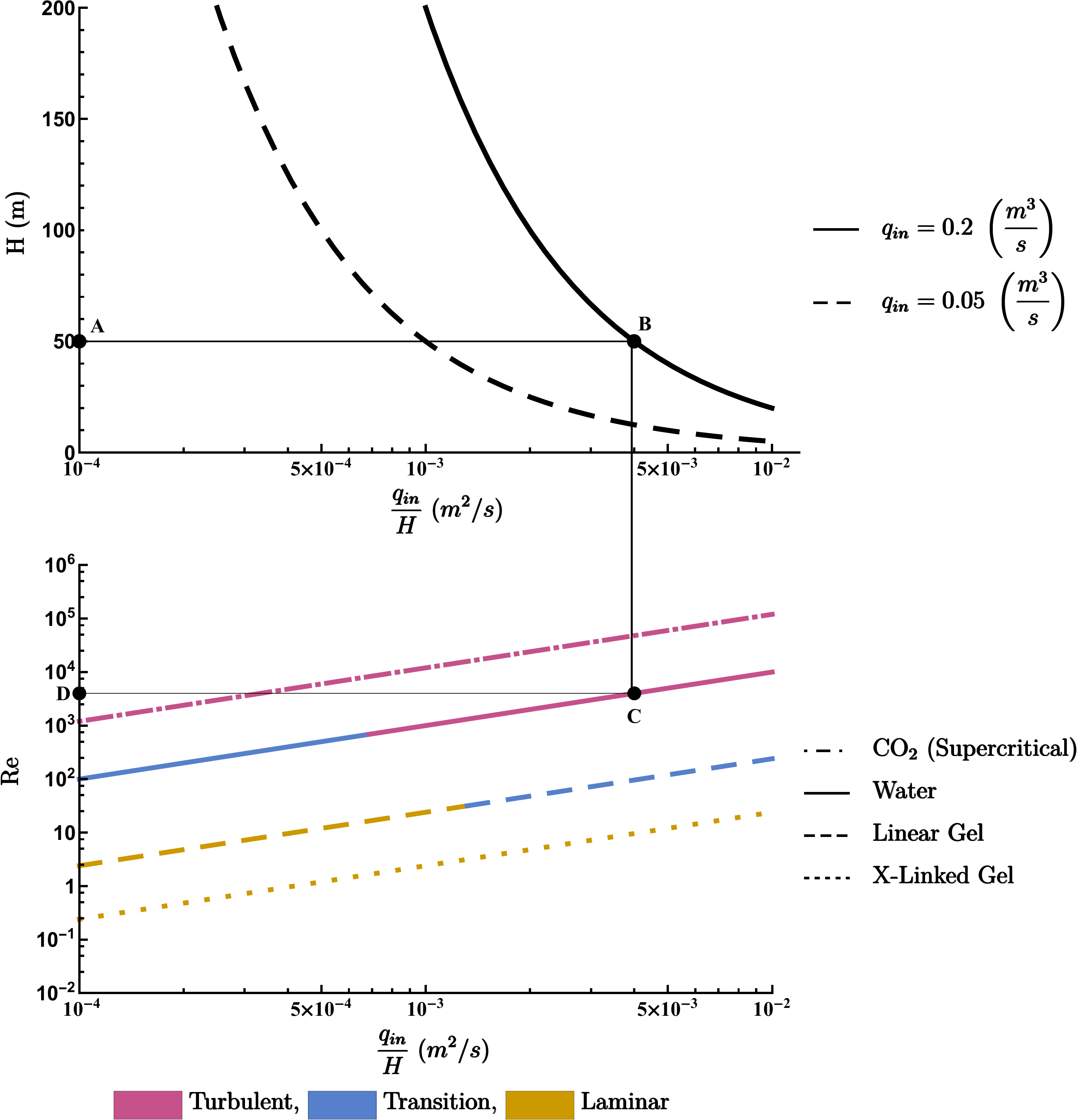}
			\caption{Flow regime variation with regimes determined by comparison of characteristic pressures.}
			\label{Figure 10}
		\end{figure}

		By increasing the Reynolds number above 500, turbulent bursts begin to occur inside the flow \cite{emmons2012}.
		When the Reynolds number increases beyond a critical value, the whole flow is fully-developed turbulence. The transition to turbulent flow is depending on parameters like roughness and channel geometry. As a general case, some previous texts offer 12500 as the limit for Reynolds number. However, they also note that this number is an estimation and is not well defined number for all cases (e.g.~\citeNP{munson02}). Moreover, Reynolds numbers above 500 have turbulent behavior but the frequency of this turbulent behavior will change. Hence, the transition remains imprecisely defined.
		Recalling the Nordgren and turbulent flow solution for pressure:
		\begin{subequations}\label{pressure_tot}
			\begin{align}
			p_{net}^N(0,t)=2.5 &\left[\frac{\mu q_{in}^2}{H^6}\frac{G^4}{(1-\nu)^4}\right]^{\frac{1}{5}}t^{\frac{1}{5}} \label{Nordgren_solutionS} \\
			p_{net}^{T}(x,t)=0.8122~\left(\frac{kq_{in}^9\rho^3}{H^{22}}\left(\frac{G}{1-\nu}\right)^{13}\right)^{\frac{1}{16}}t^{\frac{3}{16}} &\left(1-~\frac{x}{\ell(t)}\right)^{\frac{3}{7}}\left(1+0.05497~\frac{x}{\ell(t)}\right) \label{ImplicitFORMallS}
			\end{align}
		\end{subequations}
		An alternative method for defining transition is by looking at the condition that fluid pressure from the laminar solution equates with the pressure found from our turbulent solution. Specifically, letting $P_{laminar}=P_{net}^N(0,t)$ (Eq.~\ref{Nordgren_solutionS}) and letting $P_{turb}=P_{net}^T(0,t)$ (Eq.~\ref{ImplicitFORMallS}), we define $P_{laminar}>P_{turb}$ as the laminar regime and $P_{laminar}<P_{turb}$ as the turbulent regime. Finding the transition point entails solution to
		\[P_{net_{N}}(0,t)\bigg{|}_{\text{eq.~\ref{Nordgren_solutionS}}}=P_{net}(0,t)\bigg{|}_{\text{eq.~\ref{ImplicitFORMallS}}},\]
		which simplifies to
		\begin{equation}\label{equate_pressure}
		\log\left(\frac{q_{in}}{H}\right)=3.005+\frac{1}{13}\log\left[\frac{\mu^{16}Ht}{\rho^{15}k^5}\left(\frac{1-\nu}{G}\right)\right]
		\end{equation}
		Here we can see the transition point depends not only on $q_{in}/H$ but also on $\mu, H, t, \rho, k, \nu,$ and $G$. Thus, we cannot define the transition based exclusively on $Re^{*}$. Still, for ranges of these parameters we can define transitions specifically for each of the 4 fluids. The result is shown in Fig.~\ref{Figure 10} (for range of different parameters see Table~\ref{Table 4}). Again the relevant range is between the $q_{in}=0.05$ m$^3$ s$^{-1}$ and $q_{in}=0.2$ m$^3$ s$^{-1}$ curves and note that the transition region is here defined as the range in which different choices from the typical ranges of parameters can result in either $P_{laminar}>P_{turb}$ or vice versa. Within this range, by the definition of Eq.~\ref{equate_pressure}, CO$_2$ is always turbulent and water is turbulent for nearly all relevant cases.
		\begin{table}
			\caption{Limit of variables in Eq.~\protect\ref{equate_pressure}}
			\centering \vspace{0.25 cm}
			\renewcommand{\arraystretch}{1.5}
			\begin{tabu}{>{\centering}m{2.5cm} >{\centering}m{2.5cm} >{\centering}m{2.5cm}} 
				\hline\hline
				Parameter         & Minimum          & Maximum         \\ 
				\hline 
				$H$               & $20$ m           & $200$ m         \\ 
				$t$               & $0.001$ s      & $1000$ s      \\ 
				$k$               & $0.5$ mm         & $50$ mm         \\ 
				$\nu$             & $0$              & $0.5$           \\ 
				$G$               & $1$ GPa          & $100$ GPa      \\ 
			\end{tabu} 
			\label{Table 4} 
		\end{table}  
		
		\section{Near-Tip Full Asymptotics}
		Here we start with the equation
		\begin{equation}
		-\Omega+ \frac{4\varphi+1}{4\varphi+2}(\xi \Omega)' = \left[ \Omega^{2\varphi}\left(- \Omega' \right)^{1/2}\right]'.
		\label{ode_omega_full}
		\end{equation}
		We know that $\Omega$ will be small so we scale it with power series solution as
		\[\Omega=\epsilon^{a}f+\epsilon^{a+\gamma}g+\ldots~~\text{and}~~a,~\gamma>0.\]
		The distance to the fracture tip is scaled as 
		\[\xi = 1-\epsilon^{b}\zeta+\ldots~~\text{and}~~b>0,\]
		Derivatives with respect to $\xi$ become derivatives with respect to $\zeta$ through
		\[\td{}{\xi} = -\epsilon^{-b} \td{}{\zeta},\]
		which says that the derivative is large near the fracture tip, as expected. 
		\subsection{First order analysis}
		We start by examining the leading order equations. Thus, we can rewrite Eq.~\eqref{ode_omega_full} as
		\begin{equation}
		-\epsilon^{a}f- \frac{4\varphi+1}{4\varphi+2}\epsilon^{-b}\left[(1-\epsilon^{b}\zeta) \epsilon^{a}f\right]' = -\epsilon^{-b}\left[ \left(\epsilon^{a}f\right)^{2\varphi}\left( \epsilon^{a-b}f' \right)^{1/2}\right]',
		\label{ode_omega_full_scaled}
		\end{equation}
		where the primes now indicate derivatives with respect to $\zeta$. Combining terms we find
		\begin{equation}
		-\epsilon^{a}f- \frac{4\varphi+1}{4\varphi+2}\epsilon^{a-b}f'+ \frac{4\varphi+1}{4\varphi+2}\epsilon^{a}\left[\zeta f\right]' = -\epsilon^{\left(2\varphi+\frac{1}{2}\right)a -\frac{3}{2}b}\left[f^{2\varphi}\left(f' \right)^{1/2}\right]',
		\label{ode_omega_full_scaled_simplified}
		\end{equation}
		We can see that the first and third terms scale as $\epsilon^{a}$ and are smaller than the second and fourth terms. Therefore, we equate the powers of the second and fourth terms and find
		\[b = \left(4\varphi-1\right)a,\]
		which is true for any positive $a$ so long as $\varphi>1/4$. In general GMS equation, $\varphi=(3+\alpha)/4$, and considering $\varphi>1/4$ will lead to $\alpha>-2$ which is always true; for example, in Manning equation $\alpha=1/3>-2$ which satisfy $\varphi>1/4$. Now we need to choose a positive value for $a$. Thus, if we choose $a=1$, our dominant balance will become 
		\begin{equation}
		\frac{4\varphi+1}{4\varphi+2}f' = \left[ f^{2\varphi}\left(f'\right)^{1/2}\right]'.
		\label{ode_f_dominant}
		\end{equation}
		This is the same equation as derived heuristically before, but the dependent variable $u=(1-\xi)/\epsilon^{b}$. The boundary conditions are flipped and therefore,
		\[f(\zeta=0)=0,~~~\left[f^{2\varphi}(f')^{1/2}\right]_{\zeta=0}=0.\]
		Integrating Eq.~\eqref{ode_f_dominant} and solving for it will direct us to
		\[f=\left[\sqrt{4\varphi-1}\left(\frac{4\varphi+1}{4\varphi+2}\right)\right]^{2/(4\varphi-1)} \zeta^{1/(4\varphi-1)}.\]
		We can reinsert $\zeta=(1-\xi)/\epsilon^{b}$ and find
		\[\Omega=\epsilon^{a}f=\epsilon^{b/(4\varphi-1)}f=\left[\sqrt{4\varphi-1}\left(\frac{4\varphi+1}{4\varphi+2}\right)\right]^{2/(4\varphi-1)} \left(1-\xi\right)^{1/(4\varphi-1)},\]
		which is the result we derived previously.
		
		\subsection{Second order analysis}
		Now that we have found $f$ as a function of $\zeta$, we look to the next order in $\epsilon$ to find the function $g(\zeta)$ i.e.
		\[\Omega=\epsilon f+\epsilon^{1+\gamma}g+\ldots~~\text{and}~~\gamma>0.\]
		where $\gamma$ is an unknown exponent. The coordinate $\zeta$ is still scaled as
		\[\xi = 1-\epsilon^{b}\zeta,\]
		where $b=4\varphi-1$ and $\varphi>1/4$. 
		
		We can insert these scalings into the full ode for $\Omega$, Eq.~\eqref{ode_omega_full}, to find 
		\begin{eqnarray} -\epsilon f-\epsilon^{1+\gamma}g- \frac{4\varphi+1}{4\varphi+2}\epsilon^{-b}\left[\left( 1-\epsilon^{b}\zeta \right) \left(\epsilon f+\epsilon^{1+\gamma}g\right)\right]' \hspace{3cm}\nonumber \\
		= -\epsilon^{-b}\left[ \left(\epsilon f+\epsilon^{1+\gamma}g \right)^{2\varphi}\left(\epsilon^{1-b} f'+\epsilon^{1+\gamma-b} g' \right)^{1/2}\right]'.\label{fandg}\end{eqnarray}
		Expanding out the left side we find that 
		\begin{eqnarray} -\epsilon^{b+1} f-\epsilon^{1+\gamma+b}g- \frac{4\varphi+1}{4\varphi+2}\left(\epsilon f'+\epsilon^{1+\gamma}g'\right)+\frac{4\varphi+1}{4\varphi+2}\epsilon^{b+1} \left[\zeta \left( f+\epsilon^{\gamma} g\right)\right]'\nonumber \\
		= -\left[ \left(\epsilon f+\epsilon^{1+\gamma}g \right)^{2\varphi}\left(\epsilon^{1-b} f'+\epsilon^{1+\gamma-b} g' \right)^{1/2}\right]'.\label{fandg_simp_left}\end{eqnarray}
		We now expand the right side as
		\begin{eqnarray} -\epsilon^{b+1} f-\epsilon^{1+\gamma+b}g- \frac{4\varphi+1}{4\varphi+2}\left(\epsilon f'+\epsilon^{1+\gamma}g'\right)+\frac{4\varphi+1}{4\varphi+2}\epsilon^{b+1} \left[\zeta \left( f+\epsilon^{\gamma} g\right)\right]'\nonumber \\
		= -\left[ \epsilon f^{2\varphi}\left(f'\right)^{1/2}\left(1+2\varphi\epsilon^{\gamma} \frac{g}{f} \right)\left(1+\frac{\epsilon^{\gamma}}{2} \frac{g'}{f'} \right) \right]', \label{fandg_simp_right} \end{eqnarray}
		Now the terms that are linear in $\epsilon$ group as
		\[-\frac{4\varphi+1}{4\varphi+2} f'
		= -\left[ f^{2\varphi}\left(f'\right)^{1/2}\right]',\]
		which is the same equation that we solved in the earlier section. The terms that are a slight departure from linear in $\epsilon$
		\[-\epsilon^{b+1} f- \frac{4\varphi+1}{4\varphi+2} \epsilon^{1+\gamma} g'+\frac{4\varphi+1}{4\varphi+2}\epsilon^{b+1} \left[\zeta f\right]'
		= -\epsilon^{1+\gamma}\left[ f^{2\varphi}\left(f'\right)^{1/2}\left(2\varphi \frac{g}{f} +\frac{1}{2} \frac{g'}{f'} \right)\right]'.\]
		Thus, we set $\gamma=b$ and cross out the term $\epsilon^{b+1}$ from both side of equation. For simplicity, we also do the following change of variables 
		\[\kappa= \left[\sqrt{4\varphi-1}\left(\frac{4\varphi+1}{4\varphi+2}\right)\right]^{2/(4\varphi-1)} \ands c = \frac{4\varphi+1}{4\varphi+2},\]
		which help us to define $f(\zeta) = \kappa \zeta^{1/b}$. This gives the equation 
		\[-\kappa \zeta^{1/b}-c g'+c\left[\zeta f\right]'
		= -\left[ f^{2\varphi}\left(f'\right)^{1/2}\left(2\varphi \frac{g}{f} +\frac{1}{2} \frac{g'}{f'} \right)\right]'.\]
		After calculating derivations for $f$ and replacing them back into the equation, our equation reduces to 
		\[-\kappa \zeta^{1/b}-c g'+\frac{c\kappa }{b}(1+b)\zeta^{1/b}
		= -\frac{\kappa^{b/2}}{2b^{1/2}}\left(g' +b [\zeta g]''\right).\]
		Parameter $\kappa $, that has been defined earlier, can also be explained as
		\[\kappa ^{b/2} = b^{1/2}c,\]
		thus, we can divide by $c$ and find
		\[\frac{\kappa }{c}\zeta^{1/b}+\frac{1}{2}g'-\frac{b}{2} [\zeta g]''=\frac{\kappa }{b}(1+b)\zeta^{1/b}.\]
		This is an equidimensional equation, thus, we try for a solution of the form
		\[g=A\zeta^{\lambda},\]
		which gives
		\[\frac{\kappa }{c}\zeta^{1/b}+\frac{
			\lambda}{2}A\zeta^{\lambda-1}-\frac{b}{2}A\lambda(\lambda+1)\zeta^{\lambda-1}=\frac{\kappa }{b}(1+b)\zeta^{1/b}.\]
		To match powers on each side, we must have that $\lambda=(b+1)/b$. Thus, we can solve for $A$ from
		\[\frac{
			b+1}{b}A-\frac{b+1}{b}A
		\left(
		2b+1\right)=\frac{2\kappa }{b}(1+b)-\frac{2\kappa }{c},\]
		which gives
		\[A
		=\frac{\kappa }{c(1+b)}-\frac{\kappa }{b},\]
		This solution satisfies the boundary conditions
		\[g(0)=0\ands g^{2\varphi}g'(0)=0,\]
		and, therefore, we have
		\[g(\zeta) = \kappa \left( \frac{1}{c(1+b)}-\frac{1}{b} \right) \zeta^{(b+1)/b}.\]
		
		From $g(\zeta)$ we can determine $\Omega$ as 
		\[\Omega = \epsilon \kappa \zeta^{1/b}+\epsilon^{1+b}A\zeta^{(b+1)/b}.\]
		Using the similarity variable $\xi = 1-\epsilon^{b}\zeta$, we find that 
		\[\Omega = \kappa \left(1-\xi\right)^{1/b}+\epsilon^{1+b}A\left(\frac{1-\xi}{\epsilon^{b}}\right)^{(b+1)/b}.\]
		Thus, the powers of epsilon cancel in all terms and we find
		\[\Omega = \frac{\kappa }{b} \left(1-\xi\right)^{1/b}\left[b-1+\frac{b}{c(b+1)}+\left(1-\frac{b}{c(b+1)}\right)\xi\right].\]
		We now convert back to $\alpha$. We know that $\varphi=(3+\alpha)/4$ and $b = 4\varphi-1=\alpha+2$. The pre-factor $\kappa $ is calculated as
		\[\kappa = \left[\sqrt{4\varphi-1}\left(\frac{4\varphi+1}{4\varphi+2}\right)\right]^{2/(4\varphi-1)} = \left[\sqrt{\alpha+2}\left(\frac{\alpha+4}{\alpha+5}\right)\right]^{2/(2+\alpha)}\]
		Thus,
		\[\Omega = (\alpha+2)^{-\frac{\alpha+1}{\alpha+2}} \left[\frac{\alpha+4}{\alpha+5}\right]^{2/(\alpha+2)}  \left(1-\xi\right)^{1/(\alpha+2)}\left(1+\alpha+\frac{(\alpha+2)(\alpha+5)}{(\alpha+3)(\alpha+4)}+\left(1-\frac{(\alpha+2)(\alpha+5)}{(\alpha+3)(\alpha+4)}\right)\xi\right].\]
		A typical value for $\alpha$ is $\alpha=1/3$. This gives that
		\[\Omega = \left(1-\xi\right)^{3/7}\left(1.132+0.0714~\xi\right).\]
		
		\section{Alternative scaling approach}\label{alt_scaling}
		In this chapter an alternative approach to scale the problem is presented. This method is following in the spirit of previous scaling approach done by prior authors including~\citeNP{adachidetornay2002,savitski2002,bungerdetornay2007}. As was pointed out in the paper, non-linear partial differential equation governing the maximum opening $\omega(x,t)$ and boundary conditions, which are given by 
		\begin{equation}\label{initial_before_scaling_continuityS}
		\begin{split}
		\frac{\partial\omega}{\partial t} &=-\Xi\frac{\partial}{\partial x}\left[\omega^{2\varphi}\left(-\frac{\partial\omega}{\partial x}\right)^{\frac{1}{2}}\right]\\
		x &=\ell~~\Rightarrow~~\omega(\ell,t)=0\\
		x &=\ell~~\Rightarrow~~q(\ell,t)=0\\
		x &=0~~\Rightarrow~~q(0,t)=q_{in}
		\end{split}
		\end{equation}
		The scaling begins with introduction of dimensionless opening $\tilde{\Omega}(\xi,t)$ and length $\gamma(t)$, as well as a scaled coordinate $\xi$, according to
		\begin{equation}\label{scaling_parameters}
		\begin{split}
		\omega(x,t) &=\mathcal{W}(t)\tilde{\Omega}(\xi,t)\\
		\ell(t) &=\mathcal{L}(t)\gamma (t)\\
		\xi &= \frac{x}{\ell(t)} 
		\end{split}
		\end{equation}
		where $\mathcal{W}$ and $\mathcal{L}$ represent characteristic HF width and length, to be specified later. 
		Substituting Eq.~\ref{scaling_parameters} into Eq.~\ref{initial_before_scaling_continuityS} and the accompanying boundary conditions leads to (see Section~\ref{driving_continuity_equation_no_leakoff} from supplement data for details)
		\begin{equation}\label{scaled_continuity}
		\begin{split}
		-\frac{\xi t}{\mathcal{L}}\frac{\partial \mathcal{L}}{\partial t}\frac{\partial \tilde{\Omega}}{\partial \xi}+\frac{t}{\mathcal{W}}\frac{\partial \mathcal{W}}{\partial t}\tilde{\Omega} =-\Xi \frac{\mathcal{W}^{2\varphi -\frac{1}{2}}t}{\mathcal{L}^{\frac{3}{2}}}\frac{1}{\gamma^{\frac{3}{2}}}\frac{\partial}{\partial \xi} &\left[\tilde{\Omega}^{2\varphi}\left(-\frac{\partial \tilde{\Omega}}{\partial\xi}\right)^{\frac{1}{2}}\right]\\
		\frac{\pi H\Xi\mathcal{W}^{2\varphi}}{4q_{in}}\left(\frac{\mathcal{W}}{\mathcal{L}}\right)^{\frac{1}{2}}\frac{\tilde{\Omega}^{2\varphi}}{\gamma^{\frac{1}{2}}}\left(-\frac{\partial \tilde{\Omega}}{\partial\xi}\right)^{\frac{1}{2}} \bigg{|}_{\xi=0} &=1\\
		\mathcal{W}^{2\varphi}\left(\frac{\mathcal{W}}{\mathcal{L}}\right)^{\frac{1}{2}}\frac{\tilde{\Omega}^{2\varphi}}{\gamma^{\frac{1}{2}}}\left(-\frac{\partial \tilde{\Omega}}{\partial\xi}\right)^{\frac{1}{2}} \bigg{|}_{\xi=1} &=0\\
		\tilde{\Omega}(1,t) &=0
		\end{split}
		\end{equation}
		Observing Eq.~\ref{scaled_continuity}, three dimensionless groups can be defined:
		\begin{equation}\label{dimensionlessgroups}
		\begin{split}
		\mathcal{G}_1 &=\Xi \frac{\mathcal{W}^{2\varphi -\frac{1}{2}}t}{\mathcal{L}^{\frac{3}{2}}}\\
		\mathcal{G}_2 &=\mathcal{W}^{2\varphi}\left(\frac{\mathcal{W}}{\mathcal{L}}\right)^{\frac{1}{2}}\\
		\mathcal{G}_3 &=\frac{\pi H\Xi}{4q_{in}}
		\end{split}
		\end{equation}
		Choosing the characteristic scales $\mathcal{L}$ and $\mathcal{W}$ so that $\mathcal{G}_1=1$ and $\mathcal{G}_2=1$ leads to
		\begin{equation}\label{nondimensional_param_LW}
		\begin{split}
		\mathcal{W}(t) &=\left(\Xi t\right)^{\frac{1}{4\varphi+2}}\\
		\mathcal{L}(t) &=\left(\Xi t\right)^{\frac{4\varphi+1}{4\varphi+2}}
		\end{split}
		\end{equation}
		Now by replacing the result from Eq.~\ref{nondimensional_param_LW} into Eq.~\ref{scaled_continuity} and considering the changes of variables $\gamma^\frac{3}{4\varphi-1}\Omega=\tilde{\Omega}$ and $\gamma(t)=\lambda(t)/\varrho$, where $\varrho=\left(\frac{\pi H\Xi}{4q_{in}}\right)^{\frac{4\varphi-1}{4\varphi+2}}$, leads to the final form of the scaled governing equations (compare with Eq. 15 in the paper) 
		
		\begin{subequations}\label{scaled_continuity_after_simplification_newset}
			\begin{align}
			-\xi \frac{4\varphi+1}{4\varphi+2}\frac{d \Omega}{d \xi}+\frac{\Omega}{4\varphi+2} &=-\frac{d}{d \xi}\left[\Omega^{2\varphi}\left(-\frac{d \Omega}{d\xi}\right)^{\frac{1}{2}}\right] \label{scaled_continuity_after_simplification_newset_a} \\
			\Omega^{2\varphi}\left(-\frac{d \Omega}{d\xi}\right)^{\frac{1}{2}}\bigg{|}_{\xi=1} &=0 \label{scaled_continuity_after_simplification_newset_b}\\
			\Omega(1) &=0  \label{scaled_continuity_after_simplification_newset_c}\\
			\lambda^{\frac{4\varphi+2}{4\varphi-1}}\Omega^{2\varphi}\left(-\frac{d \Omega}{d\xi}\right)^{\frac{1}{2}}\bigg{|}_{\xi=0} &=1 \label{scaled_continuity_after_simplification_newset_d}
			\end{align}
		\end{subequations}
		The scaling has therefore resulted in an ordinary differential equation (ODE) for $\Omega$, Eq.~\ref{scaled_continuity_after_simplification_newset_a}. There are two homogeneous boundary conditions (flux at tip, Eq.~\ref{scaled_continuity_after_simplification_newset_c}, and crack width at tip, Eq.~\ref{scaled_continuity_after_simplification_newset_c}), and one inhomogeneous boundary condition (flux at source, Eq.~\ref{scaled_continuity_after_simplification_newset_d}). Importantly, this ODE depends on only the scaled coordinate $\xi$ and the parameter $\varphi$. It does not depend upon the scaled length, $\lambda$, which only enters via the now-decoupled inlet condition (Eq.~\ref{scaled_continuity_after_simplification_newset_d}). Hence, a solution can be obtained firstly for $\Omega$ and then can be substituted into Eq.~\ref{scaled_continuity_after_simplification_newset_d} to obtain $\lambda$. Furthermore, the parameter $\varphi$ is determined based on an \emph{a priori} choice of fluid flow law, for example upon specification to GMS, $\varphi=5/6$. In other words, the dependence upon input parameters such as reservoir height, rock/fluid properties, and injection rate is all accounted for by the scaling.
		
		\par
		Finally, by taking the integral from both sides of the equation from $\xi$ to 1 and imposing the boundary conditions, a convenient form of the ODE for $\Omega$ (Eq.~\ref{scaled_continuity_after_simplification_newset}) is given by 
		\begin{equation}\label{scaled_continuity_after_integral}
		\begin{split}
		\int^1_\xi \Omega d\eta=-\xi \frac{4\varphi+1}{4\varphi+2}\Omega+\Omega^{2\varphi}\left(-\frac{d \Omega}{d\xi}\right)^{\frac{1}{2}}\\
		\Omega(1) =0~~,~~ \Omega^{2\varphi}\left(-\frac{d\Omega}{d\xi}\right)^{\frac{1}{2}}\bigg{|}_{\xi=1}=0\\ \lambda^{\frac{4\varphi+2}{4\varphi-1}}\Omega^{2\varphi}\left(-\frac{d \Omega}{d\xi}\right)^{\frac{1}{2}}\bigg{|}_{\xi=0}=1
		\end{split}
		\end{equation}
		Thus the scaling equation is same as what explained in the paper.
		After solving the problem, the solution that obtained through this scaling will result in 
		\begin{figure}[H]
			\centering
			\centerline{\includegraphics[height=50mm]{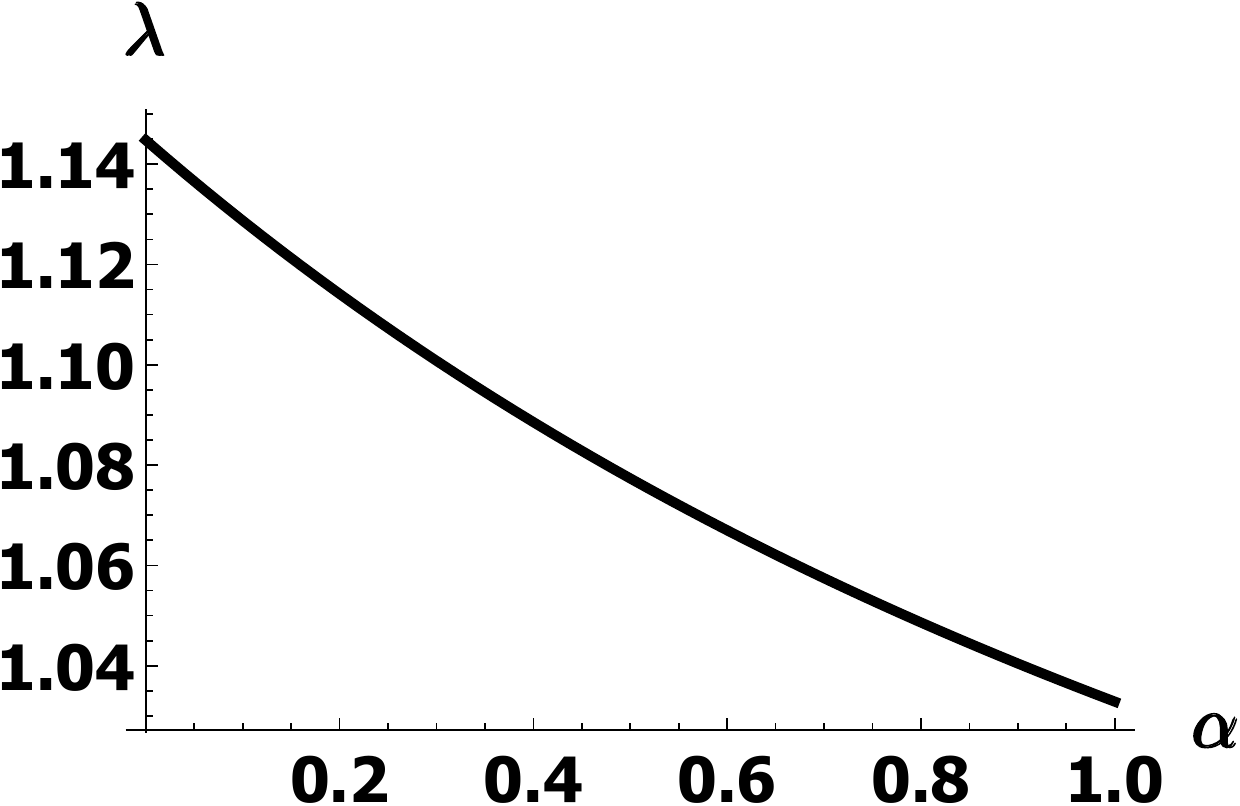}}
			\caption{Change of scaled length $\gamma$ for different values of $\alpha$.}
			\label{Figure 3}
		\end{figure}
		\subsection{Deriving Dimensionless Form of the Equations}\label{driving_continuity_equation_no_leakoff}
		After introducing the scaling in Eq.~\ref{scaling_parameters}, from application of the chain rule
		\begin{equation}\label{chain_disp}
		\frac{\partial}{\partial x}=\frac{1}{ l}\frac{\partial}{\partial \xi}=\frac{1}{\mathcal{L}\gamma}\frac{\partial}{\partial \xi}
		\end{equation}
		These two variables, $\xi$ and $t$, represent location and time, respectively. Furthermore, while $t$ is an independent variable, $\xi$ depends implicitly on time because of the time-dependence of $\ell(t)$. Hence the complete time derivative is expressed by 
		\begin{equation}
		\frac{d}{d t}= \frac{\partial}{\partial t}+\frac{\partial\xi}{\partial t}\frac{\partial}{\partial\xi}. 
		\end{equation}
		Applying to Eq.~\ref{scaling_parameters}, Eq.~\ref{initial_before_scaling_continuityS}, and the boundary conditions gives 
		\begin{equation}\label{before_involving_time}
		\begin{split}
		\frac{t}{\mathcal{W}}\frac{d(\mathcal{W}\tilde{\Omega})}{d t} =-\Xi\frac{\mathcal{W}^{2\mu-\frac{1}{2}}t}{\mathcal{L}^{\frac{3}{2}}}\frac{1}{\gamma^{\frac{3}{2}}}\frac{\partial}{\partial \xi} \left[\tilde{\Omega}^{2\mu}\left(-\frac{\partial \tilde{\Omega}}{\partial\xi}\right)^{\frac{1}{2}}\right]&\\
		\frac{\pi H\Xi\mathcal{W}^{2\varphi}}{4q_{in}}\left(\frac{\mathcal{W}}{\mathcal{L}}\right)^{\frac{1}{2}}\frac{\tilde{\Omega}^{2\varphi}}{\gamma^{\frac{1}{2}}}\left(-\frac{\partial \tilde{\Omega}}{\partial\xi}\right)^{\frac{1}{2}} \bigg{|}_{\xi=0} &=1\\
		\mathcal{W}^{2\varphi}\left(\frac{\mathcal{W}}{\mathcal{L}}\right)^{\frac{1}{2}}\frac{\tilde{\Omega}^{2\varphi}}{\gamma^{\frac{1}{2}}}\left(-\frac{\partial \tilde{\Omega}}{\partial\xi}\right)^{\frac{1}{2}} \bigg{|}_{\xi=1} &=0\\
		\tilde{\Omega}(1,t) &=0
		\end{split}
		\end{equation}
		Note we multiplied both sides by $t/\mathcal{W}$ to non-dimensionalize the equations. From the product rule and the definition of the complete time derivative
		\begin{equation}\label{LHS_expand}
		\frac{d(\mathcal{W}\tilde{\Omega})}{d t}=\left(\tilde{\Omega}\frac{d\mathcal{W}}{d t}\right)+\left(\mathcal{W}\frac{d\tilde{\Omega}}{d t}\right)=\left(\tilde{\Omega}\frac{d\mathcal{W}}{d t}\right)+\left(\mathcal{W}\left[\frac{\partial\tilde{\Omega}}{\partial t}+\frac{\partial\tilde{\Omega}}{\partial\xi}\frac{\partial\xi}{\partial t}\right]\right)
		\end{equation}
		Then using $\frac{\partial\xi}{\partial t}=-\frac{\xi}{\mathcal{L}}\frac{\partial\mathcal{L}}{\partial t}-\frac{\xi}{\gamma}\frac{\partial\gamma}{\partial t} $ leads to
		\begin{equation}\label{LHS_expand_second}
		\frac{t}{\mathcal{W}}\frac{d(\mathcal{W}\tilde{\Omega})}{d t}=t\frac{\partial\tilde{\Omega}}{\partial t}-\frac{\xi t}{\mathcal{L}}\frac{\partial\mathcal{L}}{\partial t}\frac{\partial\tilde{\Omega}}{\partial\xi}-\frac{\xi t}{\gamma}\frac{\partial\gamma}{\partial t}\frac{\partial\tilde{\Omega}}{\partial\xi}+\frac{t}{\mathcal{W}}\tilde{\Omega}\frac{d\mathcal{W}}{d t}
		\end{equation}
		It is now important to realize there are no evolution parameters appearing explicitly in the problem. This would be different if, for example, we were to consider finite leakoff; an additional term would appear in Eq.~\ref{before_involving_time} containing both a leak-off coefficient and time. However, in this limit of an impermeable rock, $\partial \tilde{\Omega}/\partial t=0$ and $d \gamma/dt=0$. Hence
		\begin{equation}\label{Limit_c_LHS}
		{\frac{t}{\mathcal{W}}\frac{d(\mathcal{W}\tilde{\Omega})}{d t}}=-\frac{\xi t}{\mathcal{L}}\frac{\partial\mathcal{L}}{\partial t}\frac{\partial\tilde{\Omega}}{\partial\xi}+\frac{t}{\mathcal{W}}\tilde{\Omega}\frac{d\mathcal{W}}{d t}
		\end{equation}
		Combining Eq.~\ref{Limit_c_LHS} and Eq.~\ref{before_involving_time} we arrive to Eq.~\ref{scaled_continuity}.
\end{document}